\documentclass[11pt]{article}
\usepackage{amsmath}
\usepackage{amsfonts}
\usepackage{amssymb}
\usepackage{epsfig}
\usepackage{times}
\usepackage{calc}
\newcommand{\p}{\partial}
\newcommand{\ol}{\overline}

\newcommand{\bs}{\boldsymbol}
\newcounter{hours}\newcounter{minutes}

\renewcommand{\l}{\newline\null}
\begin{document}
\begin{titlepage}
%
September 2001\hfill PAR-LPTHE 01/40
\vskip 4cm
{\baselineskip 17pt
\begin{center}
{\bf MASS MATRICES AND EIGENSTATES FOR SCALARS / PSEUDOSCALARS;\break
INDIRECT $\bs{CP}$ VIOLATION, MASS HIERARCHIES\break
AND  SYMMETRY BREAKING
}
\end{center}
}
\vskip .2cm
\centerline{B. Machet
     \footnote[1]{Member of `Centre National de la Recherche Scientifique'}
     \footnote[2]{E-mail: machet@lpthe.jussieu.fr}
     }
\vskip 3mm
\centerline{{\em Laboratoire de Physique Th\'eorique et Hautes \'Energies}
     \footnote[3]{LPTHE tour 16\,/\,1$^{er}\!$ \'etage,
          Universit\'e P. et M. Curie, BP 126, 4 place Jussieu,
          F-75252 Paris Cedex 05 (France)}
}
\centerline{\em Universit\'es Pierre et Marie Curie (Paris 6) et Denis
Diderot (Paris 7)}
\centerline{\em Unit\'e associ\'ee au CNRS UMR 7589}
\vskip 1cm
{\bf Abstract:}
I study indirect $CP$ violation for neutral kaons, and extend it
to large values of the $CP$-violating parameter (taken to be real).
I show how and at which condition there can exist a continuous
set of basis in which the kinetic and mass terms in the Lagrangian can
be diagonalized simultaneously. An ambiguity results
for the mass spectrum, which then depends on the basis.
In particular, for fixed (positive) $(mass)^2$ of the $CP$ eigenstates
$K^0_1, K^0_2$, and for certain ranges of values of the $CP$-violating
parameter,  a negative $(mass)^2$ can arise in the $CP$-violating basis.
Under certain conditions, even a small perturbation,
by lifting the ambiguity, can strongly alter the pattern of masses.

These investigations  extend in a natural way to indirect $CP$ violation
among a set of Higgs-like doublets.

The $C$-odd commutator $[K^0, \ol{K^0}]$, or its equivalent for Higgs
multiplets, plays an important role.
The condition for its vanishing and its consequences are among the main
concerns of this work.
\smallskip

\vfill
\null\hfil\epsffile{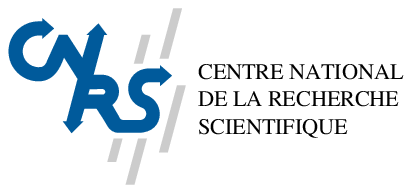}
\end{titlepage}
%
\section{Introduction}
\label{section:introduction}
%
This study concerns indirect $CP$ violation \cite{BrancoLavouraSilva}
for kaons \cite{Belusevic} and Higgs-like
doublets, when the $CP$-violating parameter is allowed to go beyond its small
customary experimental values.

It starts with the simple example of (generalized) $K_L(\theta)$ and
$K_S(\theta)$ mesons, in which $\theta$ is a real parameter
measuring indirect $CP$ violation. The decays of kaon are not
considered, such that
their mass matrix can be taken to be hermitian.
I show that there can exist a continuous set of basis, depending on $\theta$,
in which the kinetic and mass terms in the Lagrangian are both diagonal;
if so, the mass splitting depends on the basis.

The origin of the ambiguity is traced down to the contribution
of the commutator $[K^0, \ol{K^0}]$
to the mass matrix, and to the basis in which
the Lagrangian is written; in particular, the independence or not of the
vectors of the basis is important.

I investigate the vanishing of this commutator, in which case
the above-mentioned ambiguity in the mass spectrum arises.  For a pair
particle-antiparticle ($(K^0, \ol{K^0})$ or alike) considered to be
fundamental and not decaying, taking it as vanishing is legitimate;
for composite particles, like in the quark model of mesons, it is in general
untrue as soon as electroweak interactions are turned on.

I study the mass spectrum in the $CP$-violating basis  as a function of
the masses of the $CP$ eigenstates and of the $CP$-violating parameter
$\theta$.
I show that, even if, in the basis of $C\ (CP)$ eigenstates $K^0 \pm \ol{K^0}$,
all excitations are taken to be positive, the occurrence of one negative
$(mass)^2$ has nothing exceptional in the  ($C\ (CP)$
breaking) basis  $(K_L, K_S)$.

Turning on electroweak interactions makes, in general, the commutator
of composite mesons not vanish. If so,
the hamiltonian can only be diagonalized in a $CP$-violating basis,
and the masses of the $CP$-violating eigenstates $\mu_L^2$ and $\mu_S^2$
become ``physical'' observable quantities: the ambiguity in the mass
spectrum disappears.

The investigation  is straightforwardly extended
to electroweak models with more than  one Higgs doublet
\footnote{
Since the pioneering work \cite{Lee} \cite{Weinberg}, many  investigations
have been dedicated to $CP$ violation in the framework of an extended
scalar sector of the standard electroweak model (see for example
\cite{Lavoura1}\cite{Branco}\cite{LavouraSilva}\cite{BrancoLavouraSilva} and
references therein).
It has been long recognized that, since there exists no bound on the number
of Higgs doublets, it is likely to
provide an unavoidable source of $CP$ violation, in addition to the one
which comes from a complex Cabibbo-Kobayashi-Maskawa (CKM) mixing matrix
for the quarks \cite{CabibboKobayashiMaskawa}.
Much information, specially about models with two Higgs
doublets, can be found in \cite{GunionHaberKaneDawson}, and references
therein.}
.
First, it is shown that the peculiarities of the neutral kaon system can
be extended to $SU(2)_L \times U(1)$ Higgs-like doublets having opposite
transformations by $C$; then, on a more general ground, that
the same phenomenon can take place among any pair of Higgs-like doublets with
definite $C$, whatever their $C$ quantum numbers.\l
The role of the flavour singlet is emphasized.
It can in particular occur that the Higgs mass, as
it is usually defined, stays an undetermined quantity.

One is accustomed to considering both the mass difference between the $K_L$
and $K_S$ mesons and the $CP$-violating parameter $\epsilon$ as small.
When $\epsilon$ increases  unnoticed  phenomena can occur.
In particular, even a small perturbation can, under certain conditions,
induce large effects on the mass spectrum of certain pairs of particles.
While nature seems not to have pushed indirect
$CP$ violation to such extremes for pseudoscalar mesons, it
cannot be excluded  for other systems like, for example, Higgs multiplets;
no information is indeed yet available concerning
their properties  by charge conjugation.

Unlike in other works dedicated to electroweak models with several Higgs-like
doublets, no ad-hoc  potential to trigger spontaneous $CP$ breaking is
introduced; this is in particular
why such questions as ``natural flavour conservation''
\cite{GlashowWeinberg}, the presence or not of discrete symmetries
\cite{Weinberg} \cite{GlashowWeinberg}\cite{Georgi}
\cite{MendezPomarol} $\ldots$ will hardly be mentioned.

\section{How an ambiguity arises; fixing the masses of $\bs{CP}$-violating
states}
\label{section:simple}

\subsection{A unitary change of basis ($\bs T$ conserved,
 $\bs{CPT}$ violated)}
\label{subsection:unitary}

$\bullet$\ We choose the phases for the neutral pseudoscalar kaon system such
that
\begin{equation}
C(K^0) = \ol{K^0} \Rightarrow CP\ (K^0) = - \ol{K^0};
\label{eq:CP}
\end{equation}
it corresponds to $\gamma = 0$ in (\ref{eq:expand}) below.

$K^0$ and $\ol{K^0}$ are independent fields.

$\bullet$\ Let us consider the  fields
\begin{equation}
\varphi_3 = \frac{K^0 + \overline{K^0}}{\sqrt{2}},\ 
\varphi_4 = \frac{K^0 - \overline{K^0}}{\sqrt{2}}
\end{equation}
and the change of basis
\begin{equation}
\left(\begin{array}{c} \varphi_L \cr \varphi_S \end{array}\right) =
V \left(\begin{array}{c} \varphi_3 \cr \varphi_4 \end{array}\right),
\label{eq:ls}
\end{equation}
where $V$ is a unitary matrix
\begin{equation}
V = V^\dagger = V^{-1} = \left(\begin{array}{rr}   c_\theta  &  s_\theta  \cr
                         s_\theta  & -c_\theta \end{array}\right),
\label{eq:unit}
\end{equation}
and $\theta$ a real parameter (angle).

While $\varphi_3$ and $\varphi_4$ are $C$ (and $CP$) eigenstates with
opposite $C$ quantum number
\begin{equation}
\ol{\varphi_3}\equiv C(\varphi_3) = \varphi_3,
\ \ol{\varphi_4}\equiv C(\varphi_4) = - \varphi_4,
\label{eq:cp}
\end{equation}
$\varphi_L(\theta)$ and $\varphi_S(\theta)$ are, in general,
 not $C\ (CP)$ eigenstates.

At the limit of small $\theta$ they become
\footnote{This is to be compared with the usual expressions for the
$K_L$ and $K_S$ mesons \cite{Commins}
\begin{equation}
K_L = \frac{1}{(1+ \vert \epsilon_1\vert^2)^{1/2}}
            \left(\frac{K^0 + \ol{K^0}}{\sqrt{2}} +
      \epsilon_1\ \frac{K^0 - \ol{K^0}}{\sqrt{2}}\right),
K_S = \frac{1}{(1+ \vert \epsilon_2\vert^2)^{1/2}}
            \left(\frac{K^0 - \ol{K^0}}{\sqrt{2}} +
      \epsilon_2\ \frac{K^0 + \ol{K^0}}{\sqrt{2}}\right),
\end{equation}
where $\epsilon_{1,2}$ are  complex parameters.
If $CPT$ is conserved, $CP$ and $T$ violated, $\epsilon_1 = \epsilon_2$;
if $CPT$ is violated and $T$ is conserved, $\epsilon_1 = -\epsilon_2$.

The expression (\ref{eq:limit}) for small $\theta$ matches this last case. 
In addition, $P$ is conserved;  so $CP$ violation only occurs through
$C$ violation.
\label{foot:CPT}
}
\begin{equation}
\varphi_L  \longrightarrow \frac{K^0 + \ol{K^0}}{\sqrt{2}}
            + \theta\ \frac{K^0 - \ol{K^0}}{\sqrt{2}},\quad
\varphi_S \longrightarrow  \theta\ \frac{K^0 + \ol{K^0}}{\sqrt{2}}
            - \frac{K^0 - \ol{K^0}}{\sqrt{2}}.
\label{eq:limit}
\end{equation}
If $K^0$ and $\ol{K^0}$ are orthogonal and normalized to $1$
\begin{equation}
<K^0\vert K^0> = 1 = <\ol{K^0}\vert \ol{K^0}>,\quad <K^0\vert \ol{K^0}>=0
= <\ol{K^0}\vert K^0>,
\label{eq:normK}
\end{equation}
one has the relations
\begin{eqnarray}
&&<\varphi_3 \vert \varphi_3> =1 = <\varphi_4 \vert \varphi_4>,\quad
<\varphi_3 \vert \varphi_4> = 0 = <\varphi_4 \vert \varphi_3>,\cr
&& <\varphi_S \vert \varphi_S> = 1 = <\varphi_L \vert \varphi_L>,\quad
<\varphi_S \vert \varphi_L> = 0 = <\varphi_L \vert \varphi_S>.
\label{eq:nK1}
\end{eqnarray}
Because of the transformation properties (\ref{eq:cp}) of
$\varphi_3$ and $\varphi_4$ by $C$, and from our choice of phase (see
(\ref{eq:expand}) below with $\gamma=0$), we have, for operators (see for
example \cite{Pokorsky})
\begin{equation}
\ol{K^0} = C(K^0)C^{-1} = {K^0}^\dagger,
\label{eq:anti}
\end{equation}
and for the fields in the Lagrangian
\begin{equation}
\ol{K^0} = {K^0}^\ast,
\label{eq:cp2}
\end{equation}
such that (\ref{eq:cp}) rewrites
\begin{equation}
\varphi_3^\ast = \varphi_3, \quad \varphi_4^\ast = -\varphi_4;
\label{eq:real}
\end{equation}
$\varphi_3$ can be considered to be purely real, and $\varphi_4$  purely
imaginary.

With the chosen conventions, there is  equivalence between the
two notations $\ol\varphi$ (charge conjugate) and $\varphi^\ast$ (complex
conjugate); however the ``$\ast$'' notation
is useful to keep trace of which fields are independent.

$\bullet$\ The kinetic terms (${\cal L}_{kin}$) are trivially diagonal
in the three basis
$(K^0,\ol{K^0}), (\varphi_3,\varphi_4)$ and
$(\varphi_L(\theta),\varphi_S(\theta))$ since
\footnote{In particular, no $C$-odd commutator occurs here as it will in
the mass terms (see below and appendix \ref{appendix:indep}).}
\begin{equation}
{\varphi_L^\ast(\theta)}\varphi_L(\theta) + {\varphi_S^\ast(\theta)}\varphi_S(\theta)
= \varphi_4^\ast \varphi_4 + \varphi_3^\ast\varphi_3
= {K^0}^\ast K^0 + \ol{K^0}^\ast\ol{K^0}.
\label{eq:kin}
\end{equation}
They can be written in the three equivalent forms
\footnote{$K^0$ and $\ol{K^0}$ being independent, their kinetic terms are
distinct.}

\vbox{
\begin{eqnarray}
{\cal L}_{kin} &=& \frac{1}{2}\left(\p_\mu {K^0}^\ast \p^\mu {K^0}
               + \p_\mu{\ol{K^0}}^\ast \p^\mu\ol{K^0}\right)\cr
&=& \frac{1}{2}\left(\p_\mu{\varphi_3}^\ast \p^\mu\varphi_3
                    +\p_\mu{\varphi_4}^\ast \p^\mu\varphi_4\right)\cr
&=& \frac{1}{2}\left(\p_\mu{\varphi_L}^\ast\p^\mu \varphi_L +
\p_\mu{\varphi_S}^\ast\p^\mu \varphi_S\right).
\label{eq:Lkin}
\end{eqnarray}
}

$\bullet$\ Like $K^0$ and $\ol{K^0}$, $\varphi_3$ and $\varphi_4$, $\varphi_L$
and $\varphi_S$, are considered as independent fields.

Accordingly, introduce, in the $\varphi_L(\theta),\varphi_S(\theta)$ basis,
the (hermitian) mass terms
\footnote
{Consider a complex scalar field $\phi = \phi_1 + i\phi_2$,
where $\phi_1$ and $\phi_2$ are real and independent. Writing
${\cal L}_m = -m {\phi}^\ast\phi$ corresponds to a mass term for $\phi$ alone,
which re-expresses in terms of $\phi_1$ and $\phi_2$ as
${\cal L}_m= -m(\phi_1^2 + \phi_2^2 +i[\phi_1,\phi_2])$; a commutator
arises; if instead it is written
\begin{equation}
{\cal L}_m = -(m/2) ({\phi}^\ast\phi + {\phi}{\phi}^\ast),
\label{eq:Lsym}
\end{equation}
which corresponds to a mass term for $\phi$ and an identical one for
${\phi}^\ast$,  it rewrites
${\cal L}_m = -(m/2)(\phi_1^2 + \phi_2^2)$ and no commutator arises.
$[\varphi_1,\varphi_2]$ is not present when $\phi$ and
${\phi}^\ast$ are treated as two independent fields, like
$\phi_1$ and $\phi_2$.\l
$\varphi_L$, $\varphi_L^\ast$, $\varphi_S$ and
$\varphi_S^\ast$ are not independent. It is emphasized
in (\ref{eq:depend}) and (\ref{eq:genkk}), where charge conjugates can be
replaced by complex conjugates. This is why the Lagrangian (\ref{eq:Lm1})
should not be symmetrized like in (\ref{eq:Lsym}) with respect to fields
and their $\ast$ conjugates, and why  a commutator arises after using the
$C$ properties of $\varphi_3$ and $\varphi_4$ (this also traduces the
non-independence of $\varphi_3$ and $\varphi_3^\ast$, $\varphi_4$
and $\varphi_4^\ast$).
\label{foot:complex}
}
\begin{equation}
{\cal L}_m = -\frac{1}{2}\left(
\mu_S^2 {\varphi_S^\ast(\theta)}\varphi_S(\theta)
 +\mu_L^2 {\varphi_L^\ast(\theta)}\varphi_L(\theta)\right).
\label{eq:Lm1}
\end{equation}
$\mu_L^2$ and $\mu_S^2$ are considered as fixed.

${\cal L}={\cal L}_{kin} + {\cal L}_m$, given by (\ref{eq:Lkin})
(\ref{eq:Lm1}), is {\em a priori} a suitable hermitian Lagrangian
to describe one $\varphi_L$ and one $\varphi_S$ neutral meson, considered
to be, like $K^0$ and $\ol{K^0}$, two independent fields,  with
$(mass)^2$ $\mu_L^2$ and $\mu_S^2$.

In the $(\varphi_3, \varphi_4,)$ basis,
${\cal L}_m$ rewrites

\vbox{
\begin{eqnarray}
{\cal L}_m = & & -\frac{1}{2}\left(
{\varphi_3}^\ast\varphi_3 \left(\mu_S^2\sin^2\theta +
\mu_L^2\cos^2\theta \right)
             +{\varphi_4}^\ast\varphi_4 \left(\mu_L^2\sin^2\theta +
\mu_S^2\cos^2\theta \right)\right. \cr
& &\left.
+ ({\varphi_3}^\ast\varphi_4 + {\varphi_4}^\ast\varphi_3)
      \left(-\mu_S^2 +
\mu_L^2\right)\sin\theta\cos\theta\right).
\label{eq:Lm2}
\end{eqnarray}
}
The occurrence in (\ref{eq:Lm2}) of the term proportional to
$({\varphi_3}^\ast \varphi_4 + {\varphi_4}^\ast\varphi_3)$, which in particular
breaks (softly) the discrete symmetries \cite{Weinberg}\cite{GlashowWeinberg}
\cite{Georgi} \cite{MendezPomarol} $\varphi_3 \rightarrow -\varphi_3$ and
$\varphi_4 \rightarrow -\varphi_4$, comes from the fact that only
$\varphi_S$ and $\varphi_L$ are independent fields, while $(\varphi_S,
{\varphi_S}^\ast, \varphi_L, \varphi_L^\ast)$ are not (see footnote
\ref{foot:complex}, (\ref{eq:depend}), (\ref{eq:genkk}) and 
appendix \ref{appendix:indep}).

$({\varphi_3}^\ast \varphi_4 + {\varphi_4}^\ast\varphi_3)$ rewrites as
the commutator $[\varphi_3,\varphi_4] = \varphi_3\varphi_4 -
\varphi_4\varphi_3$ which is presumably vanishing (see below and section
\ref{section:commute});
if  so, in the $(\varphi_3,\varphi_4)$ basis,
${\cal L}_m$ can be considered as diagonal, too.
The masses of $\varphi_3$ and $\varphi_4$, which differ from
$\mu_S^2$ and $\mu_L^2$, are
\begin{equation}
\mu_3^2 = \mu_S^2\sin^2\theta + \mu_L^2\cos^2\theta,\quad
\mu_4^2 = \mu_S^2\cos^2\theta + \mu_L^2\sin^2\theta.
\label{eq:mumu}
\end{equation}
On Fig.~1 below are plotted $\mu_3^2$ (continuous line) and $\mu_4^2$
(dashed line) as functions of
$\theta$ for fixed values $\mu_S^2 = 4$ and $\mu_L^2 = 2$.

\vbox{
\begin{center}
{\em Fig.~1: $\mu_3^2$ and $\mu_4^2$ as functions of $\theta$
for $\mu_S^2 =4$ and $\mu_L^2 = 2$ }
\end{center}

\hskip 2.5cm
\epsfig{file=fig1.ps,height=8truecm,width=10truecm,angle=-90}
} 

\vskip -2cm
Shifting globally the pair of curves along the $y$ axis is akin to
changing $\mu_3^2
+ \mu_4^2 = \mu_L^2 + \mu_S^2$; it is then easy to see that, even for, say,
$\mu_L^2 <0$, there can exist domains for which both $\mu_3^2 $ and
$\mu_4^2$ are positive (see also section \ref{subsection:ssbu}). 

(\ref{eq:mumu}) can be inverted into
\begin{equation}
\mu_S^2 =
\frac{\mu_4^2\cos^2\theta - \mu_3^2\sin^2\theta}{\cos^2\theta -
\sin^2\theta},\quad
\mu_L^2 =
\frac{\mu_3^2\cos^2\theta - \mu_4^2\sin^2\theta}{\cos^2\theta -
\sin^2\theta}.
\label{eq:mumu2}
\end{equation}
One has
\begin{equation}
\mu_3^2 + \mu_4^2 = \mu_S^2 + \mu_L^2,\quad
\frac{\mu_3^2 - \mu_4^2}{\mu_L^2 - \mu_S^2} =
{\cos 2\theta}.
\label{eq:hier}
\end{equation}
On Fig.~2 below is plotted $(\mu_4^2- \mu_3^2)/(\mu_S^2 - \mu_L^2)$ as a
function of $\theta$.

\vbox{
\begin{center}
{\em Fig.~2: $(\mu_4^2-\mu_3^2)/(\mu_S^2 - \mu_L^2)$ as a function of $\theta$}
\end{center}

\hskip 2.5cm
\epsfig{file=fig2.ps,height=8truecm,width=10truecm,angle=-90}
} 

\vskip -2cm
Even for $\mu_L^2 \not= \mu_S^2$,
$\mu_3^2 $ and $\mu_4^2$ become degenerate for $\cos 2\theta = 0$.

For $\theta \rightarrow 0$,
\begin{equation}
\mu_4^2 \rightarrow \mu_S^2 - \theta^2 (\mu_S^2 - \mu_L^2),\quad
\mu_3^2 \rightarrow \mu_L^2 + \theta^2 (\mu_S^2 - \mu_L^2).
\label{eq:limu}
\end{equation}

$\bullet$\ Phenomena appear more clearly if one writes the mass Lagrangian
(\ref{eq:Lm1})
in the basis of independent states $(K^0,\ol{K^0})$, and also in the
$(\varphi_3,  \varphi_4)$ basis
%

\begin{eqnarray}
{\cal L}_m &\equiv& -\frac{1}{2}
\left( \begin{array}{cc} {\varphi_L}^\ast & {\varphi_S}^\ast\end{array}\right)
\left( \begin{array}{cc} \mu_L^2 & 0 \cr
                           0 & \mu_S^2 \end{array}\right)
\left( \begin{array}{c} \varphi_L \cr \varphi_S \end{array}\right)\cr
&=& -\frac{1}{4}\left[
(\mu_L^2 + \mu_S^2)({K^0}^\ast{K^0} + \ol{K^0}^\ast\ol{K^0})
+\cos 2\theta (\mu_L^2 - \mu_S^2)
\left({K^0}^\ast\ol{K^0} + \ol{K^0}^\ast K^0\right)\right.\cr
&& \left.\hskip 7cm
- \sin 2\theta (\mu_L^2 - \mu_S^2)
({K^0}^\ast K^0 - \ol{K^0}^\ast\ol{K^0})\right]\cr
&=& -\frac{1}{4}\left(\begin{array}{cc} {K^0}^\ast & \ol{K^0}^\ast\end{array}\right)
\left[\left(\begin{array}{cc}
    \mu_L^2 + \mu_S^2 & (\mu_L^2 - \mu_S^2)\cos 2\theta \cr
 (\mu_L^2 - \mu_S^2)\cos 2\theta & \mu_L^2 + \mu_S^2
 \end{array}\right)\right.\cr
&& \left.\hskip 7cm -(\mu_L^2 - \mu_S^2)\sin 2\theta
 \left(\begin{array}{cc} 1 & 0 \cr
                        0 & -1 \end{array}\right)\right]
\left(\begin{array}{c} {K^0} \cr
                       \ol{K^0} \end{array}\right)\cr
&=& -\frac{1}{2}\left(\begin{array}{cc}
            \varphi_3^\ast & {\varphi_4}^\ast\end{array}\right)
\left[\left(\begin{array}{cc} \mu_L^2\cos^2\theta + \mu_S^2\sin^2\theta & 0 \cr
     0 & \mu_S^2\cos^2\theta + \mu_L^2 \sin^2\theta
\end{array}\right)\right.\cr
&& \left.\hskip 7cm -\frac{1}{2}(\mu_L^2 - \mu_S^2){\sin 2\theta}\left(\begin{array}{cc} 0 & 1 \cr
                         1 & 0 \end{array}\right) \right]
\left(\begin{array}{c} \varphi_3 \cr
                       \varphi_4 \end{array}\right).\cr
&&
\label{eq:lm3}
\end{eqnarray}

%
The eigenvalues of the total mass matrix $M$ ({\em i.e.} the terms inside
the brackets $[\ ]$ in the last two lines of (\ref{eq:lm3})) are
$\mu_S^2$ and $\mu_L^2$, as is conspicuous from its
definition and the first line of (\ref{eq:lm3});
$M$ has been split in two parts
$M = M_1 + (\mu_L^2 - \mu_S^2)M_2\sin 2\theta$;
the last one triggers indirect $C\ (CP)$ violation
\footnote{Notice that $M_2$ has always one negative $(mass)^2$ eigenvalue.}
.

As seen in  (\ref{eq:lm3}), weak interactions are at the
origin of two phenomena, through two types of terms, both proportional to
$(\mu_L^2 -\mu_S^2)$:\l
- oscillations between the two states $K^0$ and $\ol{K^0}$, which have
degenerate masses $(\mu_L^2 + \mu_S^2)/2$; they are induced by
the $CP$ conserving $\vert\Delta S\vert = 2$ operator
$\ol{K^0}^\ast K^0 + {K^0}^\ast \ol{K^0} =
K_0^2 + \ol{K_0}^2 \equiv \varphi_3^2 + \varphi_4^2 \equiv
\varphi_L^2(\theta) + \varphi_S^2(\theta)$, with a coefficient proportional to
$\cos 2\theta$; they generate the mass splitting between the two $CP$
eigenstates $\varphi_3$ and $\varphi_4$
\footnote{They are equivalent to the customary $K^0-\ol{K^0}$
transitions triggered by weak interactions when no complex phase is present.}
;
they share similarities with Majorana mass terms which are
traditionally introduced for fermions;

- additional transitions between $\varphi_3$ and $\varphi_4$, giving rise to
indirect $C\ (CP)$ violation; they are induced by the term proportional to 
$\varphi_3^\ast\varphi_4 + \varphi_4^\ast\varphi_3$, which also rewrites
as the $CP$-odd term  $\ol{K^0}^\ast \ol{K^0} - {K^0}^\ast K^0$,
or as the commutator
$[K_0, \ol{K_0}]\equiv[\varphi_4,\varphi_3]
  \equiv [\varphi_L(\theta),\varphi_S(\theta)]$; its coefficient is
proportional to $\sin 2\theta$.

For small values of $\theta$, the second transitions correspond
to a very small energy {\em with respect to the $\phi_L-\phi_S$
mass difference}. $\mu_3^2$ and $\mu_4^2$ are very close to $\mu_S^2$ and
$\mu_L^2$.
For large values of $\theta$, $\mu_3^2$ and $\mu_4^2$
can be very different from $\mu_L^2$ and $\mu_S^2$.
 
$\bullet$\ In addition to $(\mu_L^2 + \mu_S^2)/2$ and $(\mu_L^2 -
\mu_S^2)\cos 2\theta$, (\ref{eq:lm3}) exhibits a third energy scale
\begin{equation}
\kappa^2(\theta) =\vert(\mu_L^2 - \mu_S^2)\sin 2\theta\vert.
\label{eq:kappa1}
\end{equation}
$\bullet$\ Whatever be
$\theta$, the eigenvalues of $M_1$ are $\mu_3^2$ and $\mu_4^2$ given by
(\ref{eq:mumu}) and  its eigenvectors are $\varphi_3$ and
$\varphi_4$ (identical, up to a phase, to their own antiparticles, like
Majorana fermions); a variation of $\theta$ does not change the eigenvectors,
which stay $CP$-eigenstates; $CP$ conservation
is thus associated to a $U(1)$ symmetry with phase $\theta$.

Any choice for $\theta$ breaks the above $U(1)$ symmetry;  when
$\sin 2\theta \not=0$, it  also breaks indirect $C\ (CP)$ invariance,
by adding a non-vanishing contribution to the mass matrix proportional to
$M_2$; the eigenvalues of $M$ become $\mu_S^2$ and $\mu_L^2$,
and its eigenvectors $\varphi_L(\theta)$ and $\varphi_S(\theta)$.
Then, to continue the comparison with fermion masses,
the mass eigenstates switch to states which are not
their own antiparticles, like Dirac fermions.

 When $\theta$ goes from
$0$ to $\pi/4$, indirect $C\ (CP)$ violation goes from $0$ (in which case
the mass
eigenstates are $C\ (CP)$ eigenstates) to its maximal possibility, where
$\varphi_S(\pi/4)=\ol{K^0},\varphi_L(\pi/4)={K^0}$.

The case $\theta = \pm \pi/4$ corresponds to
$\ol{\varphi_S(\pm\pi/4)} = \pm \varphi_L(\pm\pi/4)$. Then, $CPT$ requires
the two
particles to have the same mass. For instance, let us consider the case
 $\theta =+\pi/4$, for which $\varphi_L(+\pi/4) = K_0$,
$\varphi_S(+\pi/4)=\ol{K^0}$.
In the second line of (\ref{eq:lm3}), the coefficient $\cos 2\theta$ of the
$\vert \Delta S\vert = 2$ terms vanishes. As far as the $CP$ violating
term is concerned, if the commutator
$[\ol{K^0},K^0] \equiv[\varphi_3,\varphi_4]$
can be taken as identically vanishing, the condition
$\mu_L^2 = \mu_S^2$ is not needed
{\em stricto sensu} to get the identical masses required by $CPT$
for $K^0$ and $\ol{K^0}$; they are then equal to
$(\mu_L^2 + \mu_S^2)/2$, in agreement with (\ref{eq:mumu}).
If the commutator cannot be taken as vanishing, then the condition
$\mu_L^2 = \mu_S^2$ is required, which also yields  $\mu_3^2 = \mu_4^2$.

We shall come back  further in the paper to the singular points, like the
one considered above, which  correspond to $\tan^2\theta = 1$.

$\bullet$\ Unlike $K^0$ and $\ol{K^0}$,
$\varphi_L$, $\ol{\varphi_L}$, $\varphi_S$ and $\ol{\varphi_S}$
are not independent (see also Appendix \ref{appendix:indep}):
they satisfy for example the relation
\begin{equation}
(\cos\theta -\sin\theta)(\ol{\varphi_S(\theta)} + \varphi_L(\theta))
= (\cos\theta + \sin\theta)(\ol{\varphi_L(\theta)} - \varphi_S(\theta)),
\label{eq:depend}
\end{equation}
and the general equations
\begin{equation}
\left(\begin{array}{c} \ol{\varphi_L} \cr
                       \ol{\varphi_S} \end{array}\right) =
\left( \begin{array}{rr} \cos 2\theta  &  \sin 2\theta \cr
                         \sin 2\theta  &  -\cos 2 \theta \end{array}\right)
     \left(\begin{array}{c} \varphi_L \cr
                       \varphi_S \end{array}\right).
\label{eq:genkk}
\end{equation}
They form an over-complete basis; because of the equivalence between charge
conjugation and complex conjugation, it is also the case, through
(\ref{eq:real}), for $\varphi_L, \varphi_L^\ast, \varphi_S, \varphi_S^\ast$.
 
$\bullet$\ The property that ${\cal L}_m$ is diagonal in the basis
$(\varphi_3,\varphi_4)$ of $C\ (CP)$ eigenstates rests on the implicit
assumption that the commutator $[K^0(x),\ol{K^0}(x)]$, for fields at the same
space-time point,  which occurs in $M_2$, gives a vanishing contribution
to the action
\begin{equation}
\int d^4x \left(\ol{K^0}(x) K^0(x) - K^0(x) \ol{K^0}(x)\right) =0.
\label{eq:commut}
\end{equation}
This statement is confirmed by the standard expansion of $K^0$ and $\ol{K^0}$
considered as fundamental fields ($\gamma$ is an arbitrary phase that we
have chosen equal to $0$)
\begin{eqnarray}
K^0(x) &=& \int \frac{d^3k}{(2\pi)^3 2k_0}\left(
 a(\vec k)e^{-ik.x} + b^\dagger(\vec k)e^{ik.x}\right),\cr
\ol{K^0}(x) &=& e^{-i\gamma}\int \frac{d^3l}{(2\pi)^3 2l_0}\left(
 b(\vec l)e^{-il.x} + a^\dagger(\vec l)e^{il.x}\right),
\label{eq:expand}
\end{eqnarray}
in terms of independent
creation $(a^\dagger,b^\dagger)$ and annihilation $(a,b)$ operators
satisfying the usual commutation relations
\begin{equation}
[a(\vec k),a^\dagger(\vec l)] = [b(\vec k),b^\dagger(\vec l)]
= (2\pi)^3 2k_0 \delta^3(\vec k - \vec l).
\end{equation}
We shall come again to this point in subsection \ref{subsection:KK},
and show that, instead, their commutator is likely not to vanish
when $K_0$ and $\ol{K_0}$ are taken,
like in the quark model, as composite.\l
If so, $<K^0 \vert H \vert \ol{K^0}> \not= <\ol{K^0} \vert H \vert K^0>$
($H$ being the Hamiltonian), and $CPT$ is
expected to be broken. This is in agreement with our starting formulae for
$\varphi_L$ and $\varphi_S$ (\ref{eq:ls}) (see footnote \ref{foot:CPT}).
$CP$ violation occurs here only through $C$ violation, $P$ and $T$ being
conserved.

\subsection{A non-unitary transformation ($\bs{CPT}$ conserved,
$\bs T$ violated)}
\label{subsection:nonunit}

One uses the same phase convention as in subsection \ref{subsection:unitary}, such
that the notations $\varphi^\ast$ and $\ol{\varphi}$ are equivalent.

Subsection \ref{subsection:unitary} dealt with a unitary transformation
to go from $(\varphi_3, \varphi_4)$ to $(\varphi_S, \varphi_L)$.
We have mentioned (see footnote \ref{foot:CPT}) that this choice of
$\varphi_L$ and $\varphi_S$ is akin to considering that $CP$ is violated
through a violation of $CPT$, $T$ being conserved (and $P$ too).

This is not the only possibility, and the recent measurements of CPLEAR
\cite{CPLEAR}\cite{AlvarezGaume} showed that $T$ is probably violated.
Then,  $K_L$ and $K_S$ write instead, for small $\epsilon$ \cite{Commins},
that we shall suppose hereafter to be real
\begin{equation}
K_L(\epsilon) = \frac{1}{(1+ \vert \epsilon\vert^2)^{1/2}}
            \left(\varphi_3 + \epsilon\ \varphi_4\right),
K_S(\epsilon) = \frac{1}{(1+ \vert \epsilon\vert^2)^{1/2}}
            \left(\varphi_4 + \epsilon\ \varphi_3\right);
\label{eq:eps}
\end{equation}
it is the limit, for small $\epsilon$, of the non-unitary transformation
\begin{equation}
\left( \begin{array}{cc} K_L(\epsilon) \cr K_S(\epsilon) \end{array}\right)
= \frac{1}{\sqrt{\cosh^2\epsilon + \sinh^2\epsilon}}
\ A \left( \begin{array}{cc} \varphi_3 \cr \varphi_4 \end{array}\right),
\label{eq:nonun}
\end{equation}
with
\begin{equation}
A= \left(\begin{array}{cc}  \cosh\epsilon & \sinh\epsilon \cr
                      \sinh\epsilon & \cosh\epsilon \end{array}\right).
\label{eq:A}
\end{equation}
(\ref{eq:nonun}) inverts into
\begin{equation}
\left( \begin{array}{cc} \varphi_3 \cr
        \varphi_4 \end{array}\right)
= {\sqrt{\cosh^2\epsilon + \sinh^2\epsilon}}
\left(\begin{array}{cc}  \cosh\epsilon & -\sinh\epsilon \cr
                      -\sinh\epsilon & \cosh\epsilon \end{array}\right)
 \left( \begin{array}{cc} K_L(\epsilon) \cr K_S(\epsilon) \end{array}\right).\ 
\label{eq:Ainv}
\end{equation}
If $K^0$ and $\ol{K^0}$ are orthogonal and normalized to $1$ like
in (\ref{eq:normK}), $K_L$ and $K_S$ are non-longer orthogonal and
(\ref{eq:nK1}) is replaced by
\begin{equation}
 <K_S \vert K_S> = 1 = <K_L \vert K_L>,\quad
but\quad
<K_S \vert K_L> = \tanh(2\epsilon) = <K_L \vert K_S>.
\label{eq:nK21}
\end{equation}
Like in the previous section, $K_L(\epsilon), \ol{K_L(\epsilon)},
K_S(\epsilon), \ol{K_S(\epsilon)}$ are not independent. 

Performing a non-unitary transformation can have dramatic results on the
kinetic terms; in most cases, if the latter are diagonal in a basis, there
will not be after the transformation
\footnote{In particular, when the decays of the kaons are included, their
mass matrix becomes non-hermitian; the matrix which diagonalizes it
in the usual way is non-unitary, which would spoil the diagonality
of the kinetic terms. In this case, one has, instead,  to use a bi-unitary
transformation to perform the diagonalization  \cite{AlvarezGaume};
then, like for fermions, the masses of the physical states are not
the roots of the characteristic equation of their mass matrix.}
.
Of course, the spectrum of the physical states is only readable from
diagonal kinetic and mass terms.

Performing the same analysis as in section \ref{section:simple},
we transform the massive hermitian Lagrangian
\begin{equation}
{\cal L} = \frac{1}{2}\left(
           \p_\mu{K_S^\ast(\epsilon)}\p^\mu K_S(\epsilon) +
\p_\mu{K_L^\ast(\epsilon)}\p^\mu K_L(\epsilon)
 - \mu_S^2 {K_S^\ast(\epsilon)}K_S(\epsilon) - \mu_L^2
   {K_L^\ast(\epsilon)}K_L(\epsilon) \right)
\label{eq:LL1}
\end{equation}
by the non-unitary transformation $A$.
$\cal L$ becomes
\begin{eqnarray}
{\cal L} &=& \frac{1}{2}
\Big(\p_\mu{\varphi_3}^\ast \p^\mu\varphi_3 +
                       \p_\mu{\varphi_4}^\ast \p^\mu\varphi_4
  + \frac {2\sinh\epsilon\cosh\epsilon}{\cosh^2\epsilon + \sinh^2\epsilon}
 (\p_\mu{\varphi_3}^\ast\p^\mu\varphi_4 + \p_\mu{\varphi_4}^\ast\p^\mu\varphi_3)\cr
&& \cr
&& -\ \frac{(\mu_L^2\cosh^2\epsilon +
\mu_S^2\sinh^2\epsilon)}
            {\cosh^2\epsilon + \sinh^2\epsilon}{\varphi_3}^\ast\varphi_3
- \frac{(\mu_L^2\sinh^2\epsilon +
           \mu_S^2\cosh^2\epsilon)}
           {\cosh^2\epsilon + \sinh^2\epsilon}{\varphi_4}^\ast\varphi_4\cr
&& -\ (\mu_L^2 + \mu_S^2)
              \frac{\sinh\epsilon\cosh\epsilon}
             {\cosh^2\epsilon + \sinh^2\epsilon} ({\varphi_3}^\ast\varphi_4+
          {\varphi_4}^\ast\varphi_3)
\Big).\cr
&&
\label{eq:LL2}
\end{eqnarray}
Note the presence of a non-diagonal kinetic term in (\ref{eq:LL2}).

If one uses the $C$ transformation properties of $\varphi_3$ and
$\varphi_4$ to transform into commutators
$\p_\mu{\varphi_3}^\ast\p^\mu\varphi_4
+ \p_\mu{\varphi_4}^\ast\p^\mu\varphi_3 = [\p_\mu\varphi_3,\p^\mu\varphi_4]$
and ${\varphi_3}^\ast\varphi_4+ {\varphi_4}^\ast\varphi_3 = [\varphi_3,\varphi_4]$,
when $[\p_\mu\varphi_3,\p^\mu\varphi_4]=0=[\varphi_3,\varphi_4]$
,
the mass and kinetic terms can be both diagonal in the two ``basis'', and one
has the relations between the masses:
\begin{eqnarray}
\mu_3^2 &=& \frac{\mu_L^2\ \cosh^2\epsilon + \mu_S^2\
\sinh^2\epsilon}{\cosh^2\epsilon + \sinh^2\epsilon},\cr
\mu_4^2 &=& \frac{\mu_L^2\ \sinh^2\epsilon + \mu_S^2\
\cosh^2\epsilon}{\cosh^2\epsilon + \sinh^2\epsilon}.
\label{eq:m1}
\end{eqnarray}
On Fig.~3 below are plotted $\mu_3^2$ and $\mu_4^2$ as functions of
$\epsilon$ for fixed $\mu_S^2 = 4$ and $\mu_L^2 = 2$.

\vbox{
\begin{center}
{\em Fig.~3: $\mu_4^2$ and $\mu_3^2$ as  functions of $\theta$ for
$\mu_S^2 = 4$ and $\mu_L^2 = 2$}
\end{center}

\hskip 2.5cm
\epsfig{file=fig3.ps,height=8truecm,width=10truecm,angle=-90}
} 

\vskip -2cm
Like in subsection \ref{subsection:unitary}, shifting the pair
of curves along the $y$ axis is akin to changing $\mu_3^2
+ \mu_4^2 = \mu_L^2 + \mu_S^2$; it is then easy to see that, even for, say,
$\mu_L^2 <0$, there can exist domains for which both $\mu_3^2 $ and
$\mu_4^2$ are positive (see also section \ref{subsection:ssbnu}). 

(\ref{eq:m1})  inverts into
\begin{eqnarray}
\mu_L^2 = {\mu_3^2\ \cosh^2\epsilon - \mu_4^2 \
\sinh^2\epsilon},\cr
\mu_S^2 = {\mu_4^2 \ \cosh^2\epsilon - \mu_3^2\ \sinh^2\epsilon}.
\label{eq:m2}
\end{eqnarray}
A similar  ambiguity in the mass spectrum arises as in section
\ref{subsection:unitary}.

One has
\begin{equation}
\mu_L^2 + \mu_S^2= \mu_3^2 + \mu_4^2,
\quad \frac{\mu_3^2 - \mu_4^2}{\mu_L^2 - \mu_S^2}
 = \frac{1}{\cosh 2\epsilon}.
\label{eq:spec}
\end{equation}
On Fig.~4 below is plotted $(\mu_4^2- \mu_3^2)/(\mu_S^2 - \mu_L^2)$ as a
function of $\epsilon$.

\vbox{
\begin{center}
{\em Fig.~4: $(\mu_4^2-\mu_3^2)/(\mu_S^2 - \mu_L^2)$ as a function of
$\epsilon$}
\end{center}

\hskip 2.5cm
\epsfig{file=fig4.ps,height=8truecm,width=10truecm,angle=-90}
} 

\vskip -2cm
In particular, even when $\mu_L^2 \not= \mu_S^2$,
 $\mu_3^2$ and $\mu_4^2$ become identical when $\epsilon
\rightarrow \infty$.\l
At this limit, $K_L \rightarrow K^0$, $K_S \rightarrow \ol{K^0}$;
whatever different are their  masses (and they can be, as long as these two
states are not exactly identical with the two conjugate neutral kaons),
the $CP$ eigenstates $K^0_1$ and $K^0_2$ become degenerate.
So, a large $CP$ violating parameter goes along with the degeneracy of $CP$
eigenstates.

For $\epsilon \rightarrow 0$,
\begin{equation}
\mu_4^2 \rightarrow \mu_S^2 - \epsilon^2 (\mu_S^2 - \mu_L^2),\quad
\mu_3^2 \rightarrow \mu_L^2 + \epsilon^2 (\mu_S^2 - \mu_L^2),
\label{eq:limnu}
\end{equation}
which is formally identical to (\ref{eq:limu}).

In the basis $(K^0, \ol{K^0})$ and $(\varphi_3, \varphi_4)$,
 the mass Lagrangian rewrites
\begin{eqnarray}
{\cal L}_m &=&
        -\frac{1}{4}\left(\begin{array}{cc} {K^0}^\ast & \ol{K^0}^\ast\end{array}\right)
\left[\left(\begin{array}{cc}
    \mu_L^2 + \mu_S^2 & (\mu_L^2 - \mu_S^2)/\cosh 2\epsilon \cr
 (\mu_L^2 - \mu_S^2)/\cosh 2\epsilon & \mu_L^2 + \mu_S^2
 \end{array}\right)\right.\cr
&&\left.\hskip 7cm+(\mu_L^2 + \mu_S^2)\tanh 2\epsilon \left(\begin{array}{cc} 1 & 0 \cr
                        0 & -1 \end{array}\right)\right]
\left(\begin{array}{c} {K^0} \cr
                       \ol{K^0} \end{array}\right),\cr
&=& -\frac{1}{2}\left(\begin{array}{cc} \varphi_3^\ast & {\varphi_4}^\ast\end{array}\right)
\left[\left(\begin{array}{cc} \frac{\mu_L^2\ \cosh^2\epsilon + \mu_S^2\
\sinh^2\epsilon}{\cosh^2\epsilon + \sinh^2\epsilon} & 0 \cr
                        0 & \frac{\mu_L^2\ \sinh^2\epsilon + \mu_S^2\
\cosh^2\epsilon}{\cosh^2\epsilon + \sinh^2\epsilon}
\end{array}\right)\right.\cr
&&\left.\hskip 7cm+\frac{1}{2}(\mu_L^2 + \mu_S^2){\tanh 2\epsilon}\left(\begin{array}{cc} 0 & 1 \cr
                         1 & 0 \end{array}\right) \right]
\left(\begin{array}{c} \varphi_3 \cr
                       \varphi_4 \end{array}\right).\cr
&&
\label{eq:LL4}
\end{eqnarray}
which is to be compared with (\ref{eq:lm3}).

In addition to the two energy scales $(\mu_L^2 + \mu_S^2)$ and
$(\mu_L^2 - \mu_S^2)/\cosh 2\epsilon$, the third excitation
$\kappa^2(\epsilon)$ writes now:
\begin{equation}
\kappa^2(\epsilon) =  \vert(\mu_L^2 + \mu_S^2) \tanh 2\epsilon\vert.
\label{eq:kappa2}
\end{equation}
%

\subsection{Conclusion of this section}
\label{subsection:phyK}

This section uncovered a connection between the
mass spectrum and the charge conjugation properties of scalar mesons, and
exhibited the following peculiarities -- that we shall show to be
common to more general systems --.

A $CP$-odd, potentially vanishing, commutator
${\mathfrak C} \equiv i[K^0_1,K^0_2] = i[K^0,\ol{K^0}]$ can
alter the spectrum of the theory and trigger indirect $CP$ violation;
${\mathfrak C}
= i[K_L,K_S]$ when $K_L$ and $K_S$ are deduced from $K^0_1$ and $K^0_2$ by
a unitary transformation -- $CP$ broken,  $T$ conserved --, and 
${\mathfrak C} = i(\sinh^2\epsilon + \cosh^2\epsilon)[K_L,K_S]$ when
$K_L$ and $K_S$ are deduced from $K^0_1$ and $K^0_2$ by
a non-unitary transformation -- $CP$ broken, $T$ broken, $CPT$ conserved --.

It can be looked at from two different points of view.

At the level of operators, it  induces
transitions between the two (independent) types of particles on
which it operates.

At the level of the fields in the Lagrangian, the question of its vanishing
arises, and, linked to it, the existence of an ambiguity in the mass
spectrum. 

- If it vanishes, which seems a legitimate assumption when $\varphi_3$ and
$\varphi_4$ describe fundamental non-decaying particles
-- see section {\ref{section:commute} --,
an angle $\theta$, characterizing indirect $CP$-violation,
defines a continuous set of basis in which the mass
terms and the kinetic terms in the Lagrangian can be diagonalized;
the mass splittings depend on the basis and their ratio
(\ref{eq:hier}) depend only on $\theta$;
$\mu_L^2$ and $\mu_S^2$ (or $\mu_3^2$ and $\mu_4^2$) are then
ambiguous quantities.
Whatever be $\theta$, $CP$, which cannot be explicitly broken by the
term in the Lagrangian proportional to this commutator (since it vanishes), 
can only be spontaneously broken
\footnote{
If the vanishing of $[K^0,\ol{K^0}]$ is true at the quantum level,
the two basis $(K_L, K_S)$ and $(K^0_1,K^0_2)$ are equivalent, which means
that the neutral kaons can be truly measured in two different states,
with two different mass spectra,
a $CP$ conserving state and a $CP$ violating state.
This is not a contradictory statement in the absence of electroweak
interactions since the states can only be identified through their decays.}
;

- if it does not vanish,  which in particular occurs, as we shall see
in section \ref{section:commute}, when the fields are considered as composite
and as soon as electroweak interactions are turned on,
 $CP$ invariance is explicitly broken
in the Lagrangian and the $CP$-violating basis is the only diagonal one.
The masses of the $CP$-violating eigenstates are then no longer ambiguous
and become true observable.

For real kaons, $\epsilon$ (or $\theta$) is small, of order
$10^{-3}$.  The ``masses'' of the pairs $(K^0_1, K^0_2)$ and
$(K_L, K_S)$ (if they can be defined), only differ by $\theta^2$ or
$\epsilon^2$ (see (\ref{eq:limu}) and (\ref{eq:limnu})).
Furthermore, the individual masses of $K^0_1$ and $K^0_2$ (or of $K_L$
and $K_S$) are extremely close. This makes the two sets experimentally
indistinguishable, and the discussion concerning which set of masses are
really measured purely academic. There may exist systems, however, for
which the question could be relevant.

\section{Extension to Higgs-like doublets}
\label{section:general}

The same ambiguity as the one studied in section
\ref{section:simple} occurs for more general systems, in particular
$SU(2)_L \times U(1)$ Higgs-like multiplets.

We shall study below the case of a unitary transformation, keeping
in mind that a similar discussion can be made for a  non-unitary
transformation like the one performed in subsection \ref{subsection:nonunit}.

The phase convention for charge conjugation is the same as before, such
that there is equivalence between  $\ol{\phi}$ (charge conjugation) and
$\phi^\ast$ (complex conjugation).

\subsection{Scalar multiplets}
\label{subsection:multiplets}

We deal with $SU(2)_L \times U(1)$ multiplets isomorphic to the Higgs
multiplets of the Standard Model \cite{GlashowSalamWeinberg}.

If one considers quadruplets \cite{Machet1}
\begin{equation}
\phi = (\phi^0, \phi^3, \phi^+, \phi^-)
\end{equation}
with $\phi^+ = \frac{\phi^1+i \phi^2}{\sqrt{2}}$, $\phi^- = \frac{\phi^1-i
\phi^2}{\sqrt{2}}$,
transforming by $SU(2)_L$ with generators $T^3,\  T^+\equiv T^1+iT^2,\
T^-\equiv T^1-iT^2$
according to
\begin{eqnarray}
{T}^i_L\,.\,{\phi}^j &=&
-\frac{i}{2}\left( \epsilon_{ijk} {\phi}^k + \delta_{ij} {\phi}^0 \right),\cr
{T}^i_L\,.\,{\phi}^0 &=& \frac{i}{2}\; {\phi}^i,
\label{eq:actioneven}
\end{eqnarray}
then the two complex doublets
\begin{equation}
\Phi = \left(\begin{array}{c}     \phi^+ \cr
                              \frac{\phi^3-i\phi^0}{\sqrt{2}} \end{array}\right),
\quad
\tilde\Phi = \left(\begin{array}{c}  \frac{\phi^3+i\phi^0}{\sqrt{2}} \cr
                               -\phi^- \end{array}\right)
\end{equation}
are isomorphic to the standard Higgs doublets $\Phi$ and $\tilde\Phi = -i
(\Phi^\dagger\tau_2)^T$ of the Glashow-Salam-Weinberg model
(the $\vec\tau$'s are the Pauli matrices and the superscript ``$T$'' means
``transposed'').

Those quadruplets have been explicitly constructed in \cite{Machet1} as
quark-antiquark composite fields. If $N/2$ is the number of generations of
quarks, there exist $N^2/2$ such multiplets. They can always be arranged in
such a way
that the parity of $\phi^0$ is the opposite of the parity of $\vec \phi$, and
there are consequently two types of such multiplets, $(S^0, \vec P)$ and
$(P^0, \vec S)$, ``$S$'' and ``$P$'' meaning respectively ``scalar'' and
``pseudoscalar''
\footnote{The notation ``$P$'' is also used in this work for the parity
operator, but confusion should not arise.}
.

As soon as there are more than one generation of fermions, they
can also be classified according to their transformation by
charge conjugation $C$ \cite{Machet2}; in particular,
for two generations, which we shall deal with in this work,
among the total number of eight quadruplets, there are six
with $C=+1$ and two  with $C=-1$; that those numbers are all
even corresponds to the classification according to parity mentioned above.

The law of transformation (\ref{eq:actioneven}) entails that for any two
quadruplets $\phi$ and $\phi'$, the quadratic expression
\begin{equation}
\phi\phi' = \phi^0\phi^{'0} + \vec\phi.\vec\phi'
= \Phi^T \left( \begin{array}{rr}  0 & -1 \cr
                                 1 &  1 \end{array}\right) \tilde\Phi
\label{eq:invar}
\end{equation}
is invariant by $SU(2)_L$.

\subsection{A system of two quadruplets with opposite $\bs C$ quantum numbers}
\label{subsection:opposite}

For the reader to easily  make an link with the
simple example of section \ref{section:simple}, we start by investigating the
case of two $SU(2)_L \times U(1)$ quadruplets which have opposite
transformations by $C$. They are  generalizations, within
an $SU(2)_L \times U(1)$ group structure, of the states $K^0 \pm \ol{K^0}$
used there;  the Cabibbo mixing angle \cite{CabibboKobayashiMaskawa} plays
now an important role \cite{Machet1}\cite{Machet2}.  

For the sake of simplicity, we also postpone the general demonstration to
the next subsection.

We consider the quadratic Lagrangian for two
\footnote{We shall not consider the more general case of three or more
quadruplets.}
quadruplets $\phi_3$ and $\phi_4$
\footnote{The subscripts {\tiny 3} and {\tiny 4}
are used for compatibility with the
notations used in previous works. Accordingly, $\phi_3$ is linked with
the matrix ${\mathbb D}_3$ and $\phi_4$ with
the matrix ${\mathbb D}_4$ of \cite{Machet1}; that they are respectively
symmetric and antisymmetric in flavour space is at the origin of the
transformation of the quadruplets by $C$. The states chosen in the simple
example also reflect this fact.}
of the same type
with $C=+1$ and $C=-1$ respectively
\begin{equation}
\ol{\phi_3} = \phi_3, \quad \ol{\phi_4} = -\phi_4.
\end{equation}
We suppose for example that
they are both of the $(S^0,\vec P)$ type
.

In the $\phi$ basis, let the quadratic Lagrangian be
\begin{equation}
{\cal L} = \frac{1}{2}\left(
\p_\mu \left(\begin{array}{cc} {\phi_3}^\ast& {\phi_4}^\ast\end{array}\right)
\left(\begin{array}{cc} 1 & 0 \cr
                        0 & 1 \end{array}\right)
\p^\mu\left(\begin{array}{c} \phi_3 \cr
                              \phi_4 \end{array}\right)
- \left(\begin{array}{cc} {\phi_3}^\ast& {\phi_4}^\ast\end{array}\right)
   M
\left(\begin{array}{c} \phi_3 \cr
                              \phi_4 \end{array}\right)
\right).
\label{eq:L1}
\end{equation}
The choice of the kinetic term is guided by the fact that, for composite
quadruplets of definite $C$,
\begin{equation}
\sum_{all\ quadruplets} \ol\phi \phi = \sum_{all\ quadruplets}\phi^\ast\phi
\label{eq:Inv}
\end{equation}
is diagonal in both basis of flavour and electroweak eigenstates
\cite{Machet1}\cite{Machet2}
\footnote{According to (\ref{eq:invar}), it is
$\phi\phi \equiv \phi^0\phi^0 + \phi^3\phi^3 + \phi^+\phi^-
+ \phi^-\phi^+$ which is invariant by the gauge group. For quadruplets of
given $C$, $\ol\phi\phi = \phi^\ast\phi = \pm \phi\phi$; accordingly, the sum
(\ref{eq:Inv}) includes alternate signs when expressed in terms of the
$\phi$'s alone.}
.
Let $M$ be a real symmetric mass matrix
\begin{equation}
M = \left( \begin{array}{rr}   h_3  & h  \cr
                               h  & h_4  \end{array}\right)
\end{equation}
in which all $h$'s have dimension $[mass]^2$.

Its non-diagonal elements  contribute to the Lagrangian by
$-(1/2)h ({\phi_3}^\ast \phi_4 + {\phi_4}^\ast\phi_3) = -(1/2)h
(\ol{\phi_3} \phi_4 + \ol{\phi_4}\phi_3) = -(1/2)h(\phi_3\phi_4
-\phi_4\phi_3)$, which is likely to vanish. When it does, $\phi_3$ and
$\phi_4$ are mass eigenstates with masses $h_3$ and $h_4$.
 
We perform a change of basis
\begin{equation}
\left(\begin{array}{cc} \phi_3  \cr
                        \phi_4 \end{array}\right) =
V \left(\begin{array}{cc} \phi_L  \cr
                        \phi_S \end{array}\right),
\end{equation}
with $V$ a unitary transformation ($c_\theta$ and $s_\theta$ stand
respectively for $\cos\theta$ and $\sin\theta$)
\begin{equation}
V = V^\dagger = V^{-1} = \left(\begin{array}{rr}      c_\theta  &  s_\theta  \cr
                                 s_\theta  & -c_\theta \end{array}\right)
\end{equation}
which keeps the kinetic terms diagonal.

In the $\xi$ basis, the mass matrix
\begin{equation}
V^\dagger M V = \left(\begin{array}{cc}
c_\theta^2 h_3 + s_\theta^2 h_4 +2s_\theta c_\theta h &
       (s_\theta^2 -c_\theta^2)h + s_\theta c_\theta(h_3 - h_4) \cr
(s_\theta^2 -c_\theta^2)h + s_\theta c_\theta(h_3 - h_4) &
 s_\theta^2 h_3 + c_\theta^2 h_4 -2s_\theta c_\theta h  \end{array}\right)
\label{eq:VMV1}
\end{equation}
is diagonal for
\begin{equation}
\tan 2\theta = \frac{2h}{h_3-h_4},
\label{eq:diag}
\end{equation}
and the masses of $\phi_L$ and $\phi_S$ are, then
\begin{eqnarray}
\mu_L^2 &=& c_\theta^2 h_3 + s_\theta^2 h_4 +2s_\theta c_\theta h
        = \frac{c_\theta^2 h_3 - s_\theta^2 h_4}{c_\theta^2 -
s_\theta^2},\cr
\mu_S^2 &=& c_\theta^2 h_4 + s_\theta^2 h_3 -2s_\theta c_\theta h
        = \frac{c_\theta^2 h_4 - s_\theta^2 h_3}{c_\theta^2 -
s_\theta^2},
\label{eq:mu3mu4}
\end{eqnarray}
which can be inverted into (see also (\ref{eq:mumu}))
\begin{eqnarray}
h_3 &=& c^2_\theta \mu_L^2 + s^2_\theta \mu_S^2, \cr
h_4 &=& s^2_\theta \mu_L^2 + c^2_\theta \mu_S^2.
\label{eq:h3h4}
\end{eqnarray}
One deduces from (\ref{eq:mu3mu4})
\begin{equation}
\mu_L^2 + \mu_S^2 = h_3 + h_4,
\label{eq:sum}
\end{equation}
and, like in (\ref{eq:hier})
\begin{equation}
\frac{\mu_L^2 - \mu_S^2}{h_3 - h_4} =  \frac{1}{\cos{2\theta}}.
\label{eq:rapp}
\end{equation}
One has the relation:
\begin{equation}
\left(\begin{array}{c} \ol{\phi_L} \cr
                       \ol{\phi_S} \end{array}\right) =
\left( \begin{array}{rr} \cos 2\theta  &  \sin 2\theta \cr
                         \sin 2\theta  &  -\cos 2 \theta \end{array}\right)
     \left(\begin{array}{c} \phi_L \cr
                       \phi_S \end{array}\right),
\label{eq:cc34}
\end{equation}
such that the mass splitting in the $\xi$ basis, $\mu_L^2 - \mu_S^2$
cannot diverge.

The divergence in (\ref{eq:rapp}) could only happen for $\cos 2\theta =0$.
Then, $s^2_\theta = c^2_\theta$, and $V^\dagger M V$ in (\ref{eq:VMV1})
can only be diagonal for $h_3 = h_4$; its eigenvalues are then
$h_3 \pm h \equiv h_4 \pm h$, corresponding to
$\vert\mu_L^2 - \mu_S^2\vert = 2\vert h\vert$;
a vanishing $\cos 2\theta$ also corresponds, by (\ref{eq:cc34}), to
$\ol{\phi_L} = \pm \phi_S$, and $CPT$ requires then that the two states have
the same mass. As was already mentioned in section \ref{section:simple},
this constrains $h$ to be vanishing only if the commutator $[\phi_3,\phi_4]$
is not; if it does vanish, then $h$ can be
non-vanishing. This last remark can easily be checked directly by noticing
that the combination $h(\ol{\phi_L(\pm\pi/4)}\phi_L(\pm\pi/4)
-\ol{\phi_S(\pm\pi/4)}\phi_S(\pm\pi/4))$ which occurs,
factorised by $2s_\theta c_\theta$, in the diagonal terms of the mass matrix
(\ref{eq:VMV1}) is identical to $h[\phi_3,\phi_4]$.
So, whether the commutator is vanishing or not, the bound
$\vert \mu_L^2 - \mu_S^2\vert \leq \vert 2h\vert$ for $\cos
2\theta = 0$ always exists, which shows the absence of divergence for
$\vert \mu_L^2 - \mu_S^2\vert$.

(\ref{eq:rapp}) shows that the mass splitting in the basis of states which
are not $C$ eigenstates is always larger than the one in the basis of $C$
eigenstates.

If $h_3 = h_4$, (\ref{eq:mu3mu4}) entails $\mu_L^2 =
\mu_S^2 = h_3 = h_4$, except (see Fig.~1, with the replacement
$\mu_3^2 \rightarrow h_3$ and $\mu_4^2 \rightarrow h_4$) for $\sin^2\theta =
\cos^2\theta$; in this last case, $\mu_L^2 \not = \mu_S^2$ becomes
 compatible with $h_3 = h_4$.
Reciprocally, (\ref{eq:rapp}) shows that $\mu_L^2 = \mu_S^2$ only
when $h_3 = h_4$: two non-degenerate $\phi_3$ and $\phi_4$ states
(supposing $[\phi_3,\phi_4]=0$)
cannot be rotated into degenerate ones.

For $\theta$ small, $\sin\theta \approx \theta$ is the $\epsilon$-like 
parameter describing indirect $C\ (CP)$ violation in the neutral $\phi$
system; from (\ref{eq:diag}) (\ref{eq:rapp}), one gets
\begin{equation}
\theta \approx \sin\theta = \frac{2h}{\mu_L^2 - \mu_S^2}.
\end{equation}
Thus, a knowledge of the $C\ (CP)$ violating parameter
$\theta$ and of $\mu_L^2 - \mu_S^2$ determines the non-diagonal entry
$h$ of the mass matrix;  $h_3 - h_4$ can then obtained from (\ref{eq:diag}).

The two basis $(\ol{\phi_3},\phi_3, \ol{\phi_4}, \phi_4)$ and $(\ol{\phi_L},
\phi_L, \ol{\phi_S}, \phi_S)$ are over-complete; only $(\phi_3, \phi_4)$
and $(\phi_L, \phi_S)$ are not.

\subsection{A system of two quadruplets with identical $\bs C$ quantum numbers}
\label{subsection:equal}

We consider now the case of two quadruplets (of the same type, $(S^0, \vec
P)$ or $(P^0, \vec S)$) with the same $C$, and show
that one reaches the same conclusions. We  make here the
general demonstration, which can also be used in the previous section.

Let for example $\phi_2$ and $\phi_3$ be two quadruplets with $C=+1$
\begin{equation}
\ol{\phi_2} = \phi_2,\quad \ol{\phi_3} =\phi_3,
\label{eq:fi23}
\end{equation}
transforming by $SU(2)_L$ according to (\ref{eq:actioneven}).

The quadratic Lagrangian is chosen to be
\begin{equation}
{\cal L} = \frac{1}{2}\left(
\p_\mu \left(\begin{array}{cc} {\phi_2}^\ast& {\phi_3}^\ast\end{array}\right)
\left(\begin{array}{cc} 1 & 0 \cr
                        0 & 1 \end{array}\right)
\p^\mu\left(\begin{array}{c} \phi_2 \cr
                              \phi_3 \end{array}\right)
- \left(\begin{array}{cc} {\phi_2}^\ast& {\phi_3}^\ast\end{array}\right)
   M
\left(\begin{array}{c} \phi_2 \cr
                              \phi_3 \end{array}\right)
\right),
\label{eq:L2}
\end{equation}
with 
\begin{equation}
M = \left(\begin{array}{cc}        \lambda_2  &  \rho  \cr
                                   \sigma       &   \lambda_3
                   \end{array}\right).
\end{equation}
$\lambda_2,\lambda_3, \sigma, \rho$ all have dimension $[mass]^2$.

$\cal L$ is hermitian for $\lambda_2$ and $\lambda_3$ real, 
and for $\sigma = \bar \rho$. The non-diagonal mass term
$\rho({\phi_2}^\ast\phi_3) +\sigma({\phi_3}^\ast\phi_2)$ presumably vanishes,
owing to (\ref{eq:fi23}), for $\sigma = -\rho$, and we thus choose
\begin{equation}
\rho = -\sigma = i\nu,\quad \lambda_2\ \text{and}\ \lambda_3\ \text{real},
\label{eq:cond}
\end{equation}
where $\nu$ has the dimension $[mass]^2$.
When the commutator $[\phi_2,\phi_3]$ vanishes, $\phi_2$ and $\phi_3$ are
mass eigenstates with masses $\lambda_2$ and $\lambda_3$.

One goes from the $\phi$ basis to the $\xi$ basis according to
\begin{equation}
\left(\begin{array}{cc} \phi_2  \cr
                        \phi_3 \end{array}\right) =
V \left(\begin{array}{cc} \phi_L  \cr
                        \phi_S \end{array}\right),
\end{equation}
where $V$ is a general unitary matrix
\begin{equation}
V = \left( \begin{array}{cc}    a  &  b  \cr
                                c  &  d \end{array}\right).
\end{equation}
That
\begin{equation}
V^\dagger V = \left(
\begin{array}{cc} \vert a \vert^2 + \vert c \vert^2 &
               \bar a b + \bar c d \cr
    a \bar b + c \bar d  & \vert b \vert^2 + \vert d \vert^2
\end{array}\right) = 1
\end{equation}
requires
\begin{equation}
\bar a b + \bar c d = 0 = a \bar b + c \bar d, \quad \vert a \vert^2 +
\vert c \vert^2 = 1 = \vert b\vert^2 + \vert d\vert^2.
\label{eq:di1}
\end{equation}
The mass matrix in the $\xi$ basis
\begin{equation}
V^\dagger M V = \left( \begin{array}{cc}
   \vert a \vert^2 \lambda_2 + \vert c \vert^2 \lambda_3 + a\bar c \sigma
        + \bar a c \rho  &
  \bar a b \lambda_2 + \bar a d \rho + b \bar c \sigma + \bar c d \lambda_3
\cr
  a \bar b \lambda_2 + a\bar d \sigma + \bar b c \rho + c\bar d \lambda_3 &
 \vert b \vert^2 \lambda_2 + \vert d \vert^2 \lambda_3 + b\bar d \sigma +
\bar b d \rho
  \end{array}\right) 
\label{eq:VMV}
\end{equation}
is diagonal for
\begin{eqnarray}
\bar a b \lambda_2 + \bar a d \rho + b \bar c \sigma + \bar c d \lambda_3
   &=& 0, \cr
a \bar b \lambda_2 + a\bar d \sigma + \bar b c \rho + c\bar d \lambda_3
   &=& 0,
\label{eq:di2}
\end{eqnarray}
which are two self-conjugate equations when (\ref{eq:cond}) is satisfied.
(\ref{eq:di1}), (\ref{eq:di2}) and (\ref{eq:cond})  combine into
\begin{equation}
ac(\lambda_2 - \lambda_3) + i\nu(a^2 + c^2) = 0.
\label{eq:di3}
\end{equation}
The masses of $\phi_L$ and $\phi_S$ are
\begin{eqnarray}
\mu_L^2 &=& \frac{\vert a \vert^2 \lambda_2 + \vert c \vert^2 \lambda_3
   + a\bar c \bar\rho + \bar a c \rho }{\vert a \vert^2 + \vert c \vert ^2}
= {\vert a \vert^2 \lambda_2 + \vert c \vert^2 \lambda_3
      + i\nu(\bar a c - a \bar c)},\cr
\mu_S^2 &=& \frac{\vert b \vert^2 \lambda_2 + \vert d \vert^2 \lambda_3
 + b\bar d \bar\rho + \bar b d \rho}{\vert b \vert^2 + \vert d \vert ^2}
= {\vert c \vert^2 \lambda_2 + \vert a \vert^2 \lambda_3
      - i\nu(\bar a c - a \bar c)},
\label{eq:mu2mu3}
\end{eqnarray}
where we have again made use of (\ref{eq:di1}) to write the r.h.s. of
(\ref{eq:mu2mu3}).

$\bullet$\ The first trivial solution of (\ref{eq:di3})
\begin{equation}
\lambda_2 = \lambda_3,\quad c^2 + a^2 =0
\label{eq:triv}
\end{equation}
yields also through (\ref{eq:di1}) $b^2 + d^2 = 0$. Choosing, in order that
$\Delta \equiv \det V = ad - bc$ be non-vanishing,
\begin{equation}
c= + ia,\ d= -ib,
\label{eq:cd}
\end{equation}
the masses of $\phi_L$ and $\phi_S$ become
\begin{equation}
\mu_L^2 = \lambda_2 -  \nu,\quad \mu_S^2 = \lambda_2 +  \nu.
\end{equation}
However, (\ref{eq:cd}) entails
\begin{equation}
\phi_L = \frac{\phi_2 -i \phi_3}{2a},\quad \phi_S = \frac{\phi_2 +i
\phi_3}{2b},
\end{equation}
giving
\begin{equation}
a^\ast \phi_L^\ast = b\phi_S\quad or\quad \ol{a\phi_L} = b\phi_S.
\label{eq:conj}
\end{equation}
$a\phi_L$ and $b\phi_S$, being charge conjugate, must have the same
mass, which is also the mass of $\phi_L$ and $\phi_S$; indeed, one checks
explicitly that the contribution of $\rho$ in the quadratic Lagrangian for 
$\phi_L$ and $\phi_S$ identically vanishes when one makes use of
(\ref{eq:conj}), leaving a single state ($\phi_L$ or $\phi_S$) with a mass
equal to $\lambda_2 \equiv \lambda_3$. This solution we consequently discard.

$\bullet$\ We consider a more general solution to (\ref{eq:di3}).
Without loss of generality, we take $a$ to be real. Writing
\begin{equation}
\frac{c}{a} = r = r_1 + i r_2,
\label{eq:r}
\end{equation}
(\ref{eq:di3}) yields the two equations
\begin{eqnarray}
r_1 (\lambda_2 - \lambda_3) -2\nu r_1 r_2 &=& 0,\cr
r_2 (\lambda_2 - \lambda_3) + \nu ( 1 + r_1^2 - r_2^2) &=& 0.
\label{eq:r1r2}
\end{eqnarray}
The first equation  entails that, either $r_1 = 0$ or
$r_2 = (\lambda_2 - \lambda_3)/2\nu$. The second option is easily discarded
since, plugged into the second equation of (\ref{eq:r1r2}) it yields
$\lambda_2 = \lambda_3,\ \nu=0$, which is a trivial uninteresting
solution. 

So the solution of (\ref{eq:r1r2}) is
\begin{equation}
r_1 = 0,\quad r_2 = \frac
     {\lambda_2 - \lambda_3 \pm \sqrt{(\lambda_2 - \lambda_3)^2 + 4\nu^2}}
{2\nu}.
\label{eq:solr}
\end{equation}
As, from (\ref{eq:r}), $c= ia r_2$ with $a$ real, (\ref{eq:di1}) yields
$b = idr_2$, $a^2 (1+r_2^2) = 1 = \vert d \vert^2(1 + r_2^2)$.
Parameterizing $r_2 = \tan\beta$, one gets $a^2 = \vert d\vert^2 =
c^2_\beta$
\footnote{$c_\beta$ and $s_\beta$ stand respectively for
$\cos\beta$ and $\sin\beta$.}.

The masses become
\begin{eqnarray}
\mu_L^2 &=& \lambda_2 c^2_\beta + \lambda_3 s^2_\beta
-2\nu s_\beta c_\beta = 
\frac{\lambda_2 c^2_\beta - \lambda_3 s^2_\beta}{c^2_\beta -
s^2_\beta},\cr\quad
\mu_S^2 &=& \lambda_2 s^2_\beta + \lambda_3 c^2_\beta
+2\nu s_\beta c_\beta = 
\frac{\lambda_3 c^2_\beta - \lambda_2 s^2_\beta}{c^2_\beta -
s^2_\beta},
\label{eq:mumu23}
\end{eqnarray}
where we have used the second equation of (\ref{eq:r1r2}) with $r_1=0$ to
write the last members of the r.h.s. (\ref{eq:mumu23}) can be inverted for
$\lambda_2$ and $\lambda_3$ exactly like (\ref{eq:mu3mu4}) has been inverted
into (\ref{eq:h3h4}) for $h_3$ and $h_4$.

One has the relations, analogous to (\ref{eq:sum}) and (\ref{eq:rapp}):
\begin{equation}
\mu_L^2 + \mu_S^2 = \lambda_2 + \lambda_3, \quad 
\frac{\mu_L^2 - \mu_S ^2}{\lambda_2 - \lambda_3} = \frac{1}{\cos 2\beta}.
\label{eq:massrel}
\end{equation}
The eigenstates $\phi_L$ and $\phi_S$ write
\begin{eqnarray}
\phi_L &=& \frac{1}{\Delta}(d \phi_2 -b \phi_3) =
      -\frac{1}{\Delta}\ \frac{b}{\bar c}
                     \ (\bar a \phi_2 + \bar c \phi_3)
            = c_\beta \phi_2 - is_\beta\phi_3,\cr
&&\cr
\phi_S &=& \frac{1}{\Delta}(-c \phi_2 + a \phi_3)
 = \frac{c_\beta}{d}(-is_\beta\phi_2 + c_\beta\phi_3).
\label{eq:neweigen}
\end{eqnarray}
The equivalent of (\ref{eq:cc34}) is
\begin{equation}
\left(\begin{array}{c} \ol{\phi_L} \cr
                       \ol{\phi_S} \end{array}\right) =
\left( \begin{array}{cc} \cos 2\beta  &  i({d}/{c_\beta})\sin 2\beta \cr
    \frac{i}{2}\left(1 + {c_\beta}/{\bar d}\right) \sin 2\beta  &
           ({d}/{\bar d})\cos 2 \beta \end{array}\right)
     \left(\begin{array}{c} \phi_L \cr
                       \phi_S \end{array}\right),
\label{eq:cc23}
\end{equation}
which simplifies, for $d = c_\beta$ into
\begin{equation}
\left(\begin{array}{c} \ol{\phi_L} \cr
                       \ol{\phi_S} \end{array}\right) =
\left( \begin{array}{cc} \cos 2\beta  &  i\sin 2\beta \cr
     i\sin 2\beta  & \cos 2 \beta \end{array}\right)
     \left(\begin{array}{c} \phi_L \cr
                       \phi_S \end{array}\right),
\label{eq:cc23s}
\end{equation}
showing again that the ratio of mass splittings (\ref{eq:mumu23}) cannot
diverge, since, for $\cos 2\beta =0$ the states $\phi_L$ and $\phi_S$,
connected by charge conjugation, must have the same mass.

(\ref{eq:neweigen}) shows that $\phi_L$ and $\phi_S$
 are not $C$ eigenstates;
for $\beta \not = \pm \pi/4$ they are not related to each other by charge
conjugation and their masses may thus be different.

Like in the previous section, two non-degenerate $\phi_2,\phi_3$
(supposing $[\phi_2,\phi_3]=0$) cannot be rotated into degenerate ones.
Reciprocally, $\lambda_2 = \lambda_3$ entails $\mu_L^2 = \mu_S^2$, except
when  $\epsilon \rightarrow \infty$, where $\lambda_2 = \lambda_3$ is
compatible with $\mu_L^2 \not= \mu_S^2$.

The two basis $(\ol{\phi_2},\phi_2, \ol{\phi_3}, \phi_3)$ and $(\ol{\phi_L},
\phi_L, \ol{\phi_S}, \phi_S)$ are over-complete; only $(\phi_2, \phi_3)$
and $(\phi_L,\phi_S)$ are not.

\section{Commutators; fundamental versus composite scalars}
\label{section:commute}

In all examples given, an ambiguity  arose from
introducing, in the Lagrangian, a term proportional to a $C$-odd
commutator of two scalar fields.

We shall investigate here the cases when such a  commutator vanishes or not.

\subsection{The case of $\bs{K^0}$ and $\bs{\ol{K^0}}$}
\label{subsection:KK}

In the simple example of section \ref{section:simple}, we assumed that
(\ref{eq:commut}) was true, and checked the legitimacy of this statement
when $K^0$ and $\ol{K^0}$ are fundamental fields
which can be expanded according to (\ref{eq:expand}).

Suppose now that one considers them as composite fields of the type
$\bar q_i \gamma_5 q_j$, where the $q$'s are fundamental fermions (quarks).
Up to a normalization constant, it is natural to take,
in agreement with PCAC
\begin{equation}
K^0 = \frac{i}{\rho^2} \bar d \gamma_5 s,\quad
    \ol{K^0} = \frac{i}{\rho^2} \bar s \gamma_5 d,
\label{eq:Kq}
\end{equation}
where $\rho$ is a mass scale introduced to restore the correct
dimension.
By using the standard anticommutation relations of the quark fields
$\{q_i(\vec x,t),q_j^\dagger(\vec x',t)\} = \delta^3(\vec x - \vec
x')\delta_{ij}$, it is
then straightforward to calculate the commutator
\footnote{One gets of course the standard result of Current Algebra for the
commutator of two charges.}
\begin{equation}
[K^0,\ol{K^0}](x) \equiv [K_L,K_S](x)
= \frac{1}{\rho^4} (d^\dagger d - s^\dagger
s)(x)\delta^3(\vec 0)
\label{eq:comm2}
\end{equation}
and its contribution to the action
($\nu^2 \equiv \sin 2\theta (\mu_L^2 - \mu_S^2)$ in (\ref{eq:lm3}))
\begin{equation}
\nu^2\int d^4x [K^0,\ol{K^0}](x) = \frac{\nu^2}{\rho^4}\delta^3(\vec 0)
\int dt \left(N_d(t) - N_s(t)\right)
\label{eq:Ndt}
\end{equation}
where we have defined the ``charge'' $N_d(t)$ as
\begin{equation}
N_d(t) = \int d^3x\ d^\dagger(x)d(x)
=\int d^3x\ J^0_d(x)\ \text{with}\ J^\mu_d(x) = \ol{d}(x)\gamma^\mu d(x),
\label{eq:Nd}
\end{equation}
and a similar expression for $N_s(t)$.

$N_d (t)$ and $N_s (t)$ are not conserved as soon as electroweak interactions
are turned on, since they do not conserve the number of $d$ quarks nor the
one of $s$ quarks; for example, the so called ``penguin'' diagrams induce
$d \leftrightarrow s$ transitions
\footnote{This non-conservation also occurs in the decays of the
kaons, but then, as already mentioned, their mass matrix is no longer
hermitian.}
.

So, for composite neutral kaons, the $[K^0,\ol{K^0}]$ commutator is
expected not to vanish and to explicitly break $C$ invariance in the
Lagrangian (\ref{eq:Lm2}).

It would instead vanish for a neutral particle and its antiparticle
if they are identical, like the neutral pion.

The $\delta^3(\vec 0)$ in (\ref{eq:comm2}) (\ref{eq:Ndt}) originate 
from anticommuting fermions at
the same point in space, and the question of its regularization arises
\footnote{Since  $<0\vert d^\dagger d\vert 0> = 0
=<0\vert s^\dagger s\vert 0>$, the commutator cannot be regularized by just
subtracting its vacuum expectation value.}
.
This is however outside the limits of this study since taking the kaons as
composite transforms their Lagrangian into a set of (non-renormalizable)
4-fermions operators, which can only get an eventual meaning by introducing
an ultraviolet cut-off.
What we can nevertheless say is that the singularities present in a
commutator should be less severe than the ones occurring in other
4-fermions operators. So, if the kinetic terms can be given a
signification, {\it a fortiori} the commutator also can. It then occurs
that the
``strength'' of the perturbation induced by the commutator is controlled by
$\int dt\ (N_d(t)-N_s(t))$, which can be in principle very small. 
We shall come back in section \ref{section:obs} to the fact that a small
perturbation can induce large modifications of the mass spectrum.

\subsection{General case; role of the scalar flavour singlet}
\label{subsection:gen}

Though the case may have looked academic, we saw that the commutator
of mesons considered to be fundamental and independent fields vanishes.
This is also the case in the standard electroweak model when several
fundamental Higgs-like doublets are included. But, as soon as they as
considered as
composite, the commutator gets non-vanishing contributions from
electroweak interactions.

We investigate along this line Higgs-like doublets built as composite
quark-antiquark fields like in subsection \ref{subsection:multiplets}
 \cite{Machet1}\cite{Machet2}.
In this case, it turns out that there is one among the $N^2/2$ quadruplets
which, as can be easily verified, commutes with all $N^2/4$ other multiplets of the same type and, thus,
plays a special role; it is the one called $\Phi_1 = (S^0_1,\vec P_1)$ in
\cite{Machet1}\cite{Machet2}, constructed from the unit matrix in flavour
space; it includes the scalar flavour singlet
$S^0_1 \propto(\bar u u + \bar c c + \cdots + \bar d d + \bar s s + \cdots)$ and
its three pseudoscalar partners
\footnote{It was often chosen in other works of
the author as ``the'' Higgs quadruplet (complex doublet).}
.
It satisfies the relations
\footnote{In the same way, one checks that the $(P^0,\vec S)$ quadruplet
associated with the unit matrix commutes with all $N^2/4$ quadruplets of
the $9P^0,\vec S)$ type.}
\begin{equation}
For\ all\ k\not=1,\ 
[S^0_1,S^0_k]=0; [P^3_1, P^3_k]=0; [P^+_1,P^-_k]+[P^-_1,P^+_k]=0.
\label{eq:com1k}
\end{equation}
The spectrum of the theory can then be
ambiguous; in this precise case, it is related to
the freedom to add to the electroweak Lagrangian, like in subsection
\ref{subsection:unitary}, an arbitrary mass term
proportional to the commutator of $\Phi_1$ with another composite
Higgs-like (complex) doublet $\Phi_k,\ k\not=1$.

\section{Fixing the masses of $\bs{CP}$ eigenstates}
\label{section:fixeigen}

As shown below, even when the $(mass)^2$ of the $CP$ eigenstates are fixed
to be positive, a negative $(mass)^2$ can arise in a $C\ (CP)$
violating basis.
It is thus natural to investigate this phenomenon in relation with
the Higgs mechanism for the breaking of a continuous (gauge) symmetry.

In all this section, the $CP$-odd commutator $\mathfrak C$ defined in
subsection \ref{subsection:phyK}, or its equivalent for Higgs multiplets, is
supposed to vanish, such that there is a continuous set of basis, labeled
by  the value  the $CP$-violating parameter, in which the mass matrix
can be diagonal.
%
\subsection{The case $\bs T$ conserved, $\bs{CPT}$ violated}
\label{subsection:ssbu}

In this subsection,  the analysis is performed for a unitary change of basis,
like in subsection \ref{subsection:unitary} and section \ref{section:general}.
The case of a non-unitary
transformation  like the one studied in subsection \ref{subsection:nonunit}
will be examined in subsection \ref{subsection:ssbnu}.

Since (\ref{eq:mu3mu4}) and (\ref{eq:mumu23}) are formally identical, the
discussion that we make below for two quadruplets $\phi_3$ and $\phi_4$
with opposite $C$'s also applies to quadruplets with the same
$C$ quantum number.

\subsubsection{The spectrum in the basis of $\bs{CP}$-violating eigenstates}
\label{subsubsection:ssb}

Consider two sets of $C$-eigenstates $J=0$ fields
$\phi_3$ and $\phi_4$,
each of them being stable by a continuous (gauge) symmetry group $\cal G$,
and such that the  quadratic forms $\phi_3^2$, $\phi_4^2$,
$\phi_3\phi_4$ and $\phi_4\phi_3$ (the last two being identical by
commutativity)  are $\cal G$-invariant.

Suppose for example that their charge conjugates $\ol{\phi_3}$ and
$\ol{\phi_4}$
satisfy $\ol{\phi_3} = \phi_3$ and $\ol{\phi_4} = -\phi_4$ (an analogous
demonstration can be made if $\phi_4$ is $C$-even too).

For each set, a $\cal G$-invariant hermitian mass term can be written,
corresponding respectively to the positive $(mass)^2$ $\mu_3^2$ and
$\mu_4^2$, and the Lagrangian $L$ is:
\begin{equation}
L = \frac{1}{2}\left(
\p_\mu {\phi_3}^\ast \p^\mu\phi_3 + \p_\mu {\phi_4}^\ast \p^\mu\phi_4
- \mu_3^2 {\phi_3}^\ast\phi_3 - \mu_4^2 {\phi_4}^\ast\phi_4
    \right).
\label{eq:Lu}
\end{equation}
In the basis
\begin{equation}
\phi_L(\theta) = \phi_3\cos\theta + \phi_4\sin\theta,\quad
\phi_S(\theta) = \phi_3\sin\theta - \phi_4\cos\theta.
\label{eq:eigv}
\end{equation}
$L$ rewrites
\begin{equation}
L = L_{kin} -\frac{1}{2} 
\left( \begin{array}{cc} \phi^\ast_L & \phi^\ast_S \end{array}\right)
{\mathbb M}
\left( \begin{array}{c} \phi_L \cr \phi_S \end{array}\right)
\end{equation}
with
\begin{equation}
{\mathbb M} =
\left( \begin{array}{cc}
{c_\theta^2 \mu_3^2 + s_\theta^2 \mu_4^2} &
         \frac{1}{2}(\mu_3^2 - \mu_4^2) \sin 2\theta \cr
\frac{1}{2}(\mu_3^2 - \mu_4^2) \sin 2\theta &
  {c_\theta^2 \mu_4^2 + s_\theta^2 \mu_3^2}
\end{array}\right).
\label{eq:Lm5}
\end{equation}
The multiplets $\phi_L(\theta)$ and $\phi_S(\theta)$ are also stable
by $\cal G$ but violate $C$ indirectly.

Consider now $\cal L$ = $L + l_m$ with $l_m$ hermitian and $C$-odd given by
(it is the ``commutator'' term)
\begin{equation}
l_m = \frac{1}{4}(\mu_3^2 - \mu_4^2) \tan 2\theta({\phi_3}^\ast\phi_4
+ {\phi_4}^\ast\phi_3);
\label{eq:lmu}
\end{equation}
$l_m$ has both effects of canceling the $\phi_L-\phi_S$ transitions of
(\ref{eq:Lm5}) and to
shift the diagonal mass terms for $\phi_L$ and $\phi_S$, as is seen when
expressing it in the $(\phi_L,\phi_S)$ basis
\begin{equation}
l_m = \frac{1}{4}(\mu_3^2 - \mu_4^2) \tan 2\theta \left(
2s_\theta c_\theta (\phi^\ast_L\phi_L - \phi^\ast_S\phi_S)
-(c^2_\theta - s^2_\theta)(\phi^\ast_L\phi_S + \phi^\ast_S\phi_L)\right).
\label{eq:lmu2}
\end{equation}
If one writes the total mass Lagrangian 
\begin{equation}
{\cal L}_m = -\frac{1}{2}\left(\begin{array}{cc} {\phi_3}^\ast &
{\phi_4}^\ast\end{array}
\right) {\cal M}
\left(\begin{array}{c} \phi_3 \cr \phi_4 \end{array}\right),
\label{eq:Lmass}
\end{equation}
$l_m$ transforms in particular the total mass matrix from $M$ to $\cal M$ with
\begin{equation}
M= \left(\begin{array}{cc} \mu_3^2 & 0 \cr
                             0     & \mu_4^2 \end{array}\right),\quad
{\cal M} = M -\frac{1}{2}(\mu_3^2 - \mu_4^2)\tan 2\theta
 \left(\begin{array}{cc} 0 & 1 \cr
                         1 & 0 \end{array}\right).
\end{equation}
The eigenvalues of $\cal M$ are $\mu_L^2$ and $\mu_S^2$
given by (\ref{eq:mumu2});
they correspond to the eigenvectors (\ref{eq:eigv}).
 
$\cal L$, which is the same as (\ref{eq:Lm2}), is diagonal in the complex
basis $(\phi_L, \phi_S)$;
it rewrites
\begin{equation}
{\cal L} = \frac{1}{2}\left(
\p_\mu {\phi_L^\ast(\theta)} \p^\mu\phi_L(\theta)
+ \p_\mu {\phi_S^\ast(\theta)} \p^\mu\phi_S(\theta)
- \mu_L^2 {\phi_L^\ast(\theta)}\phi_L(\theta)
- \mu_S^2 {\phi_S^\ast(\theta)}\phi_S(\theta)
    \right).
\label{eq:Luu}
\end{equation}
In all cases  below, one among $(\mu_L^2,\mu_S^2)$ becomes negative

\vbox{
\begin{eqnarray}
&& For \ \ 1<\tan^2\theta< \mu_4^2/\mu_3^2,\quad \mu_S^2<0\ and\ \mu_L^2>0,\cr
&& For \ \ 1<\tan^2\theta< \mu_3^2/\mu_4^2,\quad \mu_S^2>0\ and\ \mu_L^2<0,\cr
&& For \ \ \mu_3^2/\mu_4^2<\tan^2\theta< 1,\quad \mu_S^2>0\ and\ \mu_L^2<0,\cr
&& For \ \ \mu_4^2/\mu_3^2<\tan^2\theta< 1,\quad \mu_S^2<0\ and\ \mu_L^2>0;
\label{eq:cases}
\end{eqnarray}
}

this is summarized on Fig.~5 (the dashed curve corresponds to $\mu_S^2$
and the continuous one to $\mu_L^2$), and in the formula
\begin{equation}
\mu_L^2\leq 0\ or\ \mu_S^2\leq 0 \Leftrightarrow
inf(\mu_3^2/\mu_4^2,\mu_4^2/\mu_3^2) \leq  \tan^2\theta \leq
sup(\mu_3^2/\mu_4^2,\mu_4^2/\mu_3^2).
\label{eq:ssb2}
\end{equation}

\vbox{
\begin{center}
{\em Fig.~5: An example: $\mu_L^2$ and $\mu_S^2$ as functions of $\theta$
for $\mu_3^2 =2$ and $\mu_4^2 = 4$ }
\end{center}

\hskip 2.5cm
\epsfig{file=fig5.ps,height=8truecm,width=10truecm,angle=-90}
} 

\vskip -2cm

It is straightforward to show from (\ref{eq:mumu2}) that the modulus of
the negative $(mass)^2$ is always smaller than the one of the positive
$(mass)^2$
\begin{equation}
\vert \mu^2_{<0} \vert < \mu^2_{>0}.
\label{eq:ineq}
\end{equation}

Notice from (\ref{eq:cases}) that
a very small value for $\theta$
$(\tan^2\theta \ll 1)$ in the phase where $\mu_L^2 <0$ requires the
existence of a very abrupt hierarchy $\mu_4^2/\mu_3^2 \ll 1$ or
$\mu_3^2/\mu_4^2 \ll 1$.
The other extreme case is when the two states are nearly
degenerate; then, the domain for which one of the $(mass)^2$ becomes
negative  concentrates on a very small interval $\Delta\theta
\approx 2\Delta\mu^2/\mu^2$ centered at the singular point(s)
$\theta = \pi/4 + n\pi/2$.
For exactly degenerate states, this domain shrinks to $0$
and the $U(1)$ symmetry with angle $\theta$ evoked in subsection
\ref{subsection:unitary} stays unbroken too.

In all cases, the critical values $\theta = \pm\pi/4 + n\pi/2$, equivalent to
$\sin^2\theta = \cos^2\theta$ yield singularities which must be studied
separately; they correspond to states $\phi_L(\theta),\phi_S(\theta)$
which transform into each other by charge conjugation and
which, accordingly, must have the same mass by $CPT$ (see section
\ref{section:simple} and subsection \ref{subsection:opposite}).
In these cases, we have seen that, indeed,
$\phi_L(\pm\pi/4)$ and $\phi_S(\pm\pi/4)$ can be made to have
identical masses $\mu^2 =(\mu_L^2(\pm\pi/4) + \mu_S^2(\pm\pi/4))/2$,
but that $\mu_L^2(\pm\pi/4) -\mu_S^2(\pm\pi/4)$, though its modulus stays
bounded, can be arbitrary.

Now, if one uses the identity between charge conjugation and complex
conjugation to set $\phi_3^\ast = \ol{\phi_3}=\phi_3$ and
$\phi_4^\ast = \ol{\phi_4}=-\phi_4$, $l_m$ can also be written
(see appendix \ref{subsection:appu})
\begin{equation}
l_m = \frac{1}{4} (\mu_3^2 - \mu_4^2)[\phi_3,\phi_4] \tan 2\theta
 = \frac{1}{4} (\mu_3^2 - \mu_4^2) [\phi_S,\phi_L] \tan 2\theta
=\frac{1}{4} (\mu_3^2 - \mu_4^2)[\ol{\phi},\phi] \tan 2\theta,
\label{eq:lm2}
\end{equation}
where we have introduced the independent (sets of) charge conjugate fields
$\phi$ and
$\ol{\phi}$,
\begin{equation}
\phi = \frac{\phi_3 + \phi_4}{\sqrt{2}},\quad
\ol{\phi} = \frac{\phi_3 - \phi_4}{\sqrt{2}}.
\label{eq:fifi}
\end{equation}
The last form of $l_m$ in (\ref{eq:lm2}) involves the equivalent of the
$[\ol{K^0},K^0]$ commutator of section \ref{section:simple}.
If this commutator $[\ol{\phi},\phi]=[\phi_3,\phi_4]=[\phi_L,\phi_S]$
vanishes, $L \equiv {\cal L}- l_m$ is  diagonal  is the two basis
$(\phi_3,\phi_4)$ and $(\phi_L,\phi_S)$. In the $(\phi_L,\phi_S)$ basis,
the masses are $(\mu_L^2,\mu_S^2)$.
%
\subsubsection{The Higgs mechanism}
\label{subsub:traditional}
 
Let us consider the first line of (\ref{eq:Luu}) when $\mu_L^2 < 0$
from a conservative viewpoint;
we forget in particular about the $[\phi_3,\phi_4]$ commutator such that
${\phi_L}^\ast\phi_L = c_\theta^2 \phi_3^2 + s_\theta^2 \tilde\phi_4^2$,
where $\tilde\phi_4 = i\phi_4$ is real.
To stabilize the theory, let us introduce an additional term
${L}_{4L}$ to the Lagrangian $L$
\begin{equation}
{L}_{4L} = -\frac{\lambda_L}{4} \left({\phi_L^\ast(\theta)}\phi_L(\theta)
\right)^2.
\label{eq:L4L}
\end{equation}
In the ``broken'' phase, 
\begin{equation}
<{\phi_L}^\ast(\theta)\phi_L(\theta)> =
\frac{\vert\mu_L^2\vert}{\lambda_L} \not=0,
\quad <{\phi_S}^\ast(\theta)\phi_S(\theta)> = 0,
\label{eq:VEV}
\end{equation}
such that, writing $\phi_3 \equiv (S^0_3,\vec P_3)$ and
$\tilde\phi_4 \equiv(S^0_4, \vec P_4)$ and imposing $<\vec P_3> = 0 =
<\vec P_4>$, one has  $<S^0_3> \not = 0$ and/or $<S^0_4> \not = 0$.
The gauge symmetry is spontaneously broken.

The Higgs mass squared is, like in the
standard model,  twice the modulus of
the negative mass squared in the symmetry breaking potential; hence, from
(\ref{eq:ineq}) one gets for it an absolute upper bound which is twice
the mass squared of the heaviest $J=0$ (scalar or pseudoscalar) composite
mesons which make up the other Higgs-like multiplets:
\begin{equation}
M_H^2(\theta) \equiv 2\, \vert\mu^2_{<0}(\theta)\vert  \leq 2\ sup\,(M^2_{S,P}).
\label{eq:MH}
\end{equation}
$M_H^2$ is given by
\begin{equation}
M_H^2(\theta) \equiv 2\vert\mu^2_{<0}(\theta)\vert = 
2\vert\mu_4^2 - \mu_3^2\vert
\left\vert
\frac{\tan^2\theta -\tan^2\theta_c}
{(1-\tan^2\theta)(1-\tan^2\theta_c)}\right\vert
= 2(\mu_3^2 + \mu_4^2)\frac{\tan^2\theta -
\tan^2\theta_c}{(1-\tan^2\theta)(1+\tan^2\theta_c)}
\label{eq:MH2}
\end{equation}
where
\begin{equation}
\tan^2\theta_c = \frac{\mu_3^2}{\mu_4^2}.
\label{eq:thetac}
\end{equation}
We have made $\mu_3^2 + \mu_4^2$ appear in (\ref{eq:MH2}) because it is
invariant by the change of basis (see (\ref{eq:hier})).

For a simple potential like above, the condition
$<\phi_S(\theta)> = 0$ entails $<\phi_L(\theta)> = <\phi_3>/\cos\theta$.
$\phi_L(\theta)$ and $\phi_S(\theta)$ correspond to the so-called
Georgi's basis \cite{GeorgiNanopoulos}.
The Higgs boson is the neutral component of the set $\phi_L(\theta)$ which
gets a non-vanishing vacuum expectation value; it can be a $P$ eigenstate
but it is never a $C\ (CP)$ eigenstate, nor are the three
Goldstones which are eaten by the gauge bosons to become massive
\footnote{Consequently, the $C$ properties of the massive gauge bosons
could be suspected in this framework to
be more subtle than usually considered (see also \cite{AhluwaliaKirchbach}).}.

In the case when $\mu_S^2 <0$, $M_H^2$ is given by (\ref{eq:MH2}) after
changing $\tan^2\theta$ into $1/\tan^2\theta$.
%
\subsection{The case $\bs T$ violated, $\bs{CPT}$ conserved}
\label{subsection:ssbnu}
 
We perform the same study as in section \ref{subsection:ssbu} in the case of the
non-unitary transformation of subsection \ref{subsection:nonunit}.

\subsubsection{The spectrum of states}
\label{subsubsection:specnu}

We consider the same Lagrangian $L$ as in (\ref{eq:Lu}).

In the complex basis $(\phi_L, \phi_S)$
\begin{eqnarray}
\phi_L(\epsilon) &=& \frac{1}{\sqrt{\cosh^2\epsilon + \sinh^2\epsilon}}
(\phi_3\cosh\epsilon + \phi_4\sinh\epsilon),\cr
\phi_S(\epsilon) &=& \frac{1}{\sqrt{\cosh^2\epsilon +
\sinh^2\epsilon}}(\phi_3\sinh\epsilon + \phi_4\cosh\epsilon).
\label{eq:eignu}
\end{eqnarray}
it writes
\begin{equation}
L = \frac{1}{2} \cosh^2 2\epsilon
\left(\begin{array}{cc} \phi^\ast_L & \phi^\ast_S\end{array}\right)
 \left( {\mathbb K} - {\mathbb M}\right)
\left(\begin{array}{c} \phi_L \cr \phi_S \end{array}\right)
\label{eq:Lnu}
\end{equation}
with
\begin{equation}
{\mathbb K} = p^2\left(\begin{array}{cc}
  1 & -\tanh 2\epsilon  \cr
- \tanh 2\epsilon & 1 
              \end{array}\right) 
\label{eq:Knu}
\end{equation}
and
\begin{equation}
{\mathbb M} = \left(\begin{array}{cc}
\frac{\mu_3^2\cosh^2\epsilon +\mu_4^2\sinh^2\epsilon}{\cosh^2\epsilon +
\sinh^2\epsilon}   & 
-\frac{(\mu_3^2 + \mu_4^2)}{2}\tanh 2\epsilon \cr
-\frac{(\mu_3^2 + \mu_4^2)}{2}\tanh 2\epsilon &
\frac{\mu_4^2\cosh^2\epsilon +\mu_3^2\sinh^2\epsilon}{\cosh^2\epsilon +
\sinh^2\epsilon}
              \end{array}\right)
\label{eq:Mnu}
\end{equation}

If we add to it $l_m$ and $l_\chi$ given by
\begin{eqnarray}
l_m &=& -\frac{1}{4}(\mu_3^2 + \mu_4^2) \tanh 2\epsilon({\phi_3}^\ast\phi_4
+ {\phi_4}^\ast\phi_3),\cr
l_\chi &=& \frac{1}{2}{\tanh 2\epsilon}(\p_\mu{\phi_3}^\ast\p^\mu\phi_4 +
\p_\mu{\phi_4}^\ast\p^\mu\phi_3),
\label{eq:lmnu}
\end{eqnarray}
to reconstruct ${\cal L} = L + l_m + l_\chi$ which is the analog of
(\ref{eq:LL2}),
$\cal L$ can be diagonalized in the complex  basis $(\phi_L, \phi_S)$,
and rewrites
\begin{equation}
{\cal L} = \frac{1}{2}
\left(
\p_\mu {\phi_L^\ast(\epsilon)} \p^\mu\phi_L(\epsilon)
+ \p_\mu {\phi_S^\ast(\epsilon)} \p^\mu\phi_S(\epsilon)
- \mu_L^2 {\phi_L^\ast(\epsilon)}\phi_L(\epsilon)
- \mu_S^2 {\phi_S^\ast(\epsilon)}\phi_S(\epsilon)
    \right).
\label{eq:LLnu}
\end{equation}
$\mu_L^2$ and $\mu_S^2$ are given by (\ref{eq:m2}).

The effects of $l_m$ and $l_\chi$ are conspicuous when they are rewritten
in the $(\phi_L,\phi_S)$ basis (see appendix \ref{subsection:appnu}).

\vbox{
\begin{eqnarray}
l_m &=& -\frac{1}{4}(\mu_3^2 + \mu_4^2)\sinh 2\epsilon\cr
&& \hskip 2cm \left(
 -2\sinh\epsilon\cosh\epsilon(\phi^\ast_L\phi_L + \phi^\ast_S\phi_S)
+ (\cosh^2\epsilon + \sinh^2\epsilon)(\phi^\ast_L\phi_S + \phi^\ast_S\phi_L)
\right),\cr
l_\chi &=& \frac{1}{2}\sinh 2\epsilon \cr
&& \left(
 -2\sinh\epsilon\cosh\epsilon(\p_\mu\phi^\ast_L\p^\mu\phi_L +
\p_\mu\phi^\ast_S\p^\mu\phi_S)
+ (\cosh^2\epsilon + \sinh^2\epsilon)(\p_\mu\phi^\ast_L\p^\mu\phi_S +
\p_\mu\phi^\ast_S\p^\mu\phi_L) \right).\cr
&&
\label{eq:lmchi}
\end{eqnarray}
}

Again, if one uses the identity between complex conjugation and charge
conjugation, and the $C$ properties of $\phi_3$ and $\phi_4$, $l_m$ and
$l_\chi$ become proportional to commutators ($[\phi,\ol{\phi}]$ and
$[\p_\mu\phi,\p^\mu\ol{\phi}]$, where the independent charge conjugate
fields $\phi$ and $\ol{\phi}$ are the
equivalent of (\ref{eq:fifi})) which are likely
to vanish. If this is so, $L \equiv {\cal L}-l_m -l_{\chi}$ can be
diagonalized in the two basis $(\phi_3,\phi_4)$ and $(\phi_L,\phi_S)$.
In the $(\phi_L,\phi_S)$ basis, the masses are $\mu_L^2$ and $\mu_S^2$.

On Fig.~6 are displayed $\mu_S^2$ and $\mu_L^2$ as functions of
$\epsilon$ for $\mu_3^2 =2$ and $\mu_4^2 =4$.

\vbox{
\begin{center}
{\em Fig.~6: $\mu_L^2$ and $\mu_S^2$ as functions of $\epsilon$ for
 $\mu_3^2 = 2$ and $\mu_4^2 = 4$}
\end{center}

\hskip 2.5cm
\epsfig{file=fig6.ps,height=8truecm,width=10truecm,angle=-90}
} 

\vskip -2cm
$\mu_L^2$ becomes negative for $\tanh^2\epsilon > \tanh^2\epsilon_c$
with
\begin{equation}
\tanh^2\epsilon_c = \frac{\mu_3^2}{\mu_4^2}.
\label{eq:epsc}
\end{equation}
%
\subsubsection{The Higgs mechanism}
\label{subsubsec:tradnu}

For $\mu_L^2 <0$, we are led to introduce, like previously, a ``stabilizing''
potential
given by (\ref{eq:L4L}). One still use the notation $\tilde\phi_4 =
i\phi_4$.

The Higgs mass is then $M_H^2 = 2\vert \mu_L^2\vert$ and writes
\begin{equation}
M_H^2(\epsilon) = 2 (\mu_3^2 + \mu_4^2) \cosh^2\epsilon\left\vert \frac
{\tanh^2\epsilon - \tanh^2\epsilon_c}
{1 + \tanh^2\epsilon_c}\right\vert.
\label{eq:Hnunit}
\end{equation}
$\mu_3^2 + \mu_4^2$ is again invariant by the change of basis (see
(\ref{eq:spec})).
%
\section{Mass hierarchies}
\label{section:mh}

We have paid attention to the ratios between mass splittings 
in the different basis which can diagonalize the Lagrangian when the
commutator $\mathfrak C$ vanishes.

Also  of interest are the hierarchies of
masses $\mu_4^2/\mu_3^2$ for fixed $\mu_L^2$ and $\mu_S^2$ (Figs.~7,8),
and, inversely,
the ratio $\mu_S^2/\mu_L^2$ for fixed $\mu_3^2$ and $\mu_4^2$ (Figs.~9,10).
In each of the two sets of figures, the first corresponds to a unitary change
of basis and the second to a non-unitary transformation.

The hierarchy of masses is highly dependent of the basis, or
more precisely of the parameters $\theta$ or $\epsilon$ measuring indirect
$CP$ violation.
Huge mass hierarchies (between states which are not $CP$ eigenstates)
can be brought back to small ones (by going to another basis
of $CP$ eigenstates), or vice-versa.

\vbox{
\begin{center}
{\em Fig.~7:  unitary change of basis; $\mu_4^2/\mu_3^2$ as a function of
$\theta$ for $\mu_S^2 = 4$ and $\mu_L^2 = 2$}
\end{center}
\hskip 2.5cm
\epsfig{file=fig7.ps,height=8truecm,width=10truecm,angle=-90}
} 
\vskip -2.5cm
\vbox{
\begin{center}
{\em Fig.~8: non-unitary transformation; $\mu_4^2/\mu_3^2$ as a function of
$\epsilon$ for $\mu_S^2  = 4$ and $\mu_L^2 = 2$}
\end{center}

\hskip 2.5cm
\epsfig{file=fig8.ps,height=8truecm,width=10truecm,angle=-90}
} 

\vskip -2cm
\vbox{
\begin{center}
{\em Fig.~9:  unitary change of basis; $\mu_S^2/\mu_L^2$ as a function
of $\theta$ for $\mu_3^2 = 2$ and $\mu_4^2 = 4$}
\end{center}

\hskip 4cm
\epsfig{file=fig9.ps,height=6truecm,width=10truecm,angle=-90}
} 

\vskip -2cm
\vbox{
\begin{center}
{\em Fig.~10: non-unitary transformation; $\mu_S^2/\mu_L^2$ as a function
of $\epsilon$ for $\mu_3^2  = 2$ and $\mu_4^2 = 4$}
\end{center}

\hskip 2.5cm
\epsfig{file=fig10.ps,height=8truecm,width=10truecm,angle=-90}
} 

\vskip -2cm
%

\section{Ambiguous or observable masses}
\label{section:obs}

The ambiguity that eventually arises in the mass spectrum of commuting fields
has been connected to the existence of a symmetry; for example,
in subsection \ref{subsection:unitary}, we mentioned the role of the $U(1)$
symmetry with angle $\theta$.
In general, no observable can ever be associated with an unbroken symmetry
\footnote{A typical example is colour.}
;
in the case at hand, due to the freedom to enlarge their mass matrix,
the masses of commuting states do not correspond to observable quantities.
A perturbation that breaks the symmetry in question is needed to lift this
ambiguity.

Suppose indeed that one turns on electroweak interactions. The commutator
$[\phi,\ol{\phi}]$ is then likely not to vanish;
the Lagrangian ${\cal L} = L + l_m$ (\ref{eq:Luu}) can no longer be
diagonalized in a continuous set of basis, but only in the $CP$-violating
basis $(\phi_L,\phi_S)$;
$\mu_L^2$, $\mu_S^2$ and $\theta$ become observable.
The perturbation $l_m$ (\ref{eq:lmu}) changes the mass
spectrum from $(\mu_3^2,\mu_4^2)$ to $(\mu_L^2,\mu_S^2)$, and, in
particular, alters the hierarchy pattern.

A noticeable point is that the spectrum is independent of the precise value
of the commutator. So, by tuning it, one can consider the possibility
that the perturbation $l_m$ (\ref{eq:lmu}) can be made very small.
Since it is, as we stressed in section \ref{section:commute},
 sensitive to electroweak interactions, this can be achieved,
presumably, by setting their coupling constant to small enough values.
If, in this process, the $CP$ violating parameter keeps to high enough values,
a small perturbation is likely to induce large modifications in the mass
spectrum.\l
The question is thus whether and how the $CP$-violating parameter(s)
depends on the strength of electroweak interactions.
When computed in the Standard Model (see for example \cite{Belusevic},
p. 104 and 108) $\epsilon$ turns out to be, indeed, independent
of the electroweak coupling constant. However, it depends on the mixing
angles of the CKM matrix \cite{CabibboKobayashiMaskawa}
 like $\lambda^6/\sin^2\theta_c \cos^2\theta_c$,
where $\theta_c$ is the Cabibbo angle and $\lambda$ is the small parameter of
order $\sin\theta_c$ which is introduced in the Wolfenstein parameterization
\cite{Wolfenstein} of the CKM matrix.
While the Cabibbo angle is indeed expected to go to zero when the electroweak
interactions are turned off,  one does not know explicitly how fast
\footnote{Calculating such a dependence concerns flavour physics
beyond the standard model.}
;
as far as the parameter $\lambda$, which also appears there, is concerned,
it is only a phenomenological one, the true dependence of which on the
Cabibbo angle, and hence on the electroweak coupling constant, is unknown.
The dependence of $\epsilon$ on the coupling constant is thus still
unknown, which leaves the door open for the mechanism that we just evoked.

It is thus not excluded that, even for small $\alpha$ and $g,g'$ as we
know them experimentally, the $CP$-violating parameter is large enough
such that, at the same time, $l_m$ stays a ``very small'' perturbation,
and  its effects on the mass spectrum are large,  possibly even inducing
 ``spontaneous symmetry breaking''.
If large hierarchies are observed in a ``physical'' basis of $CP$-violating
states,  we suggest that they can be ``slightly''
perturbed $CP$ eigenstates for which, in the absence of perturbation, the
hierarchies are much smaller or ever equal to $1$
\footnote{
We have seen in particular that, while $\mu_L^2 = \mu_S^2$ requires
$\mu_3^2 = \mu_4^2$, $\mu_3^2$ and $\mu_4^2$ can be degenerate even
for $\mu_L^2 \not= \mu_S^2$ (for $\theta = \pi/4 + n\pi$ or
$\vert \epsilon\vert \rightarrow \infty$)}
.
Accordingly, for such particles, the $CP$-violating parameter(s) are expected
to differ from the customary small values observed in systems like neutral
kaons.

\subsection{$\bs{CPT}$ constraint}
\label{subsection:CPT}

There is a case where the spectrum is constrained (by $CPT$)
to be unambiguous:
the one corresponding to a pair of neutral  charge-conjugate commuting states.
This corresponds to maximal indirect $CP$-violation ($\theta = \pi/4 + n\pi\
{\text or}\ \epsilon = \pm\infty$). I show below how this constraint is
recovered.

\subsubsection{Unitary change of basis}
\label{subsubsction:oscu}

Suppose that, owing to the independence and the charge conjugation
properties of $\phi_3$ and $\phi_4$,
 $\phi_3^\ast \phi_4 + \phi_4^\ast \phi_3 = [\phi_3,\phi_4]=0$; 
in the $(\phi_L,\phi_S)$ basis this rewrites (see
(\ref{eq:a3}) (\ref{eq:a4})) 
$\sin 2\theta(\phi_L^\ast \phi_L - \phi_S^\ast \phi_S) -
\cos 2\theta(\phi_L^\ast \phi_S + \phi_S^\ast \phi_L) =[\phi_S,\phi_L]= 0$.
This last identity can also be checked with the help of (\ref{eq:genkk}).
It shows that splitting the states  by the term
$(\phi_L^\ast \phi_L - \phi_S^\ast \phi_S)$
is equivalent to making them oscillate by $(\phi_L^\ast \phi_S +
\phi_S^\ast \phi_L)$,  with a proportionality factor $1/\tan 2\theta$:
\begin{equation}
\phi_L^\ast \phi_L - \phi_S^\ast \phi_S = \frac{1}{\tan 2 \theta}
         (\phi_L^\ast \phi_S + \phi_S^\ast \phi_L).
\label{eq:splos}
\end{equation}
The r.h.s. of (\ref{eq:splos}) goes to $0$ when $\theta
\rightarrow \pm \pi/4$. So, at this limit, which is the one where
$\phi_L = \pm \phi$ and $\phi_S = \pm \ol{\phi}$ are conjugate
-- see (\ref{eq:fifi}) --, one can write,
fixing $\mu_L^2$ and $\mu_S^2$ to finite values (eventually negative)

\vbox{
\begin{eqnarray}
\mu_L^2 \phi_L^\ast\phi_L + \mu_S^2 \phi_S^\ast\phi_S &=&
\mu_L^2 \phi_L^\ast\phi_L + \mu_S^2 \phi_S^\ast\phi_S
-\mu_L^2(\phi_L^\ast \phi_L - \phi_S^\ast \phi_S)
+\mu_L^2(\phi_L^\ast \phi_L - \phi_S^\ast \phi_S)\cr
&=&(\mu_L^2 + \mu_S^2)\phi_S^\ast\phi_S +
\frac{\mu_L^2}{\tan 2\theta}(\phi_L^\ast\phi_S + \phi_S^\ast\phi_L)
\rightarrow (\mu_L^2 + \mu_S^2)\phi_S^\ast\phi_S\cr
{\text or}&=& \mu_L^2 \phi_L^\ast\phi_L + \mu_S^2 \phi_S^\ast\phi_S
+\mu_S^2(\phi_L^\ast \phi_L - \phi_S^\ast \phi_S)
-\mu_S^2(\phi_L^\ast \phi_L - \phi_S^\ast \phi_S)\cr
&=&(\mu_L^2 + \mu_S^2)\phi_L^\ast\phi_L +
-\frac{\mu_S^2}{\tan 2\theta}(\phi_L^\ast\phi_S + \phi_S^\ast\phi_L)
\rightarrow (\mu_L^2 + \mu_S^2)\phi_L^\ast\phi_L.\cr
&&
\label{eq:limmass}
\end{eqnarray}
}

For independent,  commuting $\phi$ and $\ol{\phi}$ such that
$\ol{\phi} = \phi^\ast$,
(\ref{eq:limmass}) is nothing more than the trivial identity
\begin{equation}
\mu_L^2 \ol{\phi}^\ast \ol{\phi} + \mu_S^2 {\phi}^\ast \phi =
(\mu_L^2 + \mu_S^2) {\phi}^\ast\phi = (\mu_L^2 + \mu_S^2)\phi{\phi}^\ast
= \frac{\mu_L^2+ \mu_S^2}{2}(\phi^\ast \phi + \phi\phi^\ast).
\end{equation}

On the r.h.s. of (\ref{eq:limmass}) only one among the two fields
$\phi_L$, $\phi_S$ remains, and it has a mass $(\mu_S^2 + \mu_L^2)/2$
(which can be positive even if $\mu_L^2$ was negative);
the same occurs for the kinetic terms, since
$\phi_L^\ast\phi_L +\phi_S^\ast\phi_S \rightarrow 2\phi_S^\ast\phi_S$
or $\phi_L^\ast\phi_L +\phi_S^\ast\phi_S \rightarrow 2\phi_L^\ast\phi_L$ .
So, at this limit, as expected, $\phi$ and $\ol\phi$ are degenerate and
their mass is the one written above.

\subsubsection{Non-unitary transformation}
\label{subsubsction:oscnu}

When $[\phi_3,\phi_4]=0$ and since $\cosh 2\epsilon$ never vanishes,
(\ref{eq:a8}) entails
\begin{equation}
\phi_L^\ast\phi_L + \phi_S^\ast\phi_S = \frac{1}{\tanh 2\epsilon}
(\phi_L^\ast\phi_S + \phi_S^\ast\phi_L).
\label{eq:splos2}
\end{equation}
Unlike in the case of a unitary change of basis, the r.h.s. of
(\ref{eq:splos2}) never vanishes, such that, now, shifting
$\mu_L^2$ and $\mu_S^2$ always goes with a ``rotation'' of the states.

The rest of the discussion, including now the non-diagonal kinetic term,
 follows the same lines as above.
 
\subsection{The special case of the flavour singlet}
\label{subsction:flasin}

The case of the flavour singlet stays a special one since
it always commutes with other $J=0$ mesons:
either as a fundamental field, which commutes with other independent
similar fields,
or,  as shown in section \ref{section:commute}, as a composite
quark-antiquark field.

So, for  a mixture of two multiplets of the same type with definite $C$,
the first including the scalar flavour singlet, turning on electroweak
interactions is likely not to remove an ambiguity in the mass spectrum,
such that the equivalent of $\mu_L^2$ and $\mu_S^2$ keep undetermined;
this can be in particular the case of the Higgs boson mass,
$M_H^2 = 2\vert \mu^2_{<0}\vert$, which stays, in this framework,
an unobservable quantity.
\footnote{This can however be considered  unrealistic as soon as the
Higgs boson is expected to decay.}
.

Are there ways to lift this ambiguity?  A possibility
is that the states carry quantum numbers other than electroweak, 
associated with interactions which are  mis-aligned with the former
(for example flavour-diagonal ``strong'' interactions);
if a detector, being mostly sensitive to these interactions,
signs the corresponding eigenstates
\footnote{Here, the process of detection is considered to be similar to
the production mechanism. For example, charged kaons, which are considered
to be flavour eigenstates, are commonly produced by strong interactions,
which are flavour-diagonal.}
,
the process of measure  can determine which linear combinations
of electroweak $CP$ eigenstates are detected;  this eventually fixes an
orientation in the space that they span, can determine the $CP$-violating
parameters, and select, among all possibilities, a precise  mass pattern
\footnote{It is not at all guaranteed that the sole freedom in the mass
matrix studied in this work is enough to switch from the basis of
electroweak $CP$ eigenstates to the basis of outgoing states that are
detected through the other type of interactions.}
.
The latter can, accordingly, depend on the quantum numbers
that are detected.

\subsection{Another possible attitude}
\label{subsection:other}

One should not put arbitrarily aside the other reasonable attitude which
simply refutes the existence of an ambiguity. Then, that the mass spectrum of
a commuting pair be uniquely defined constrains the $CP$-violating
parameter to special values, $\theta = n\pi\ {\text or}\ \pi/2 + n\pi$
in subsection \ref{subsection:unitary}, $\epsilon = 0$ in subsection
\ref{subsection:nonunit}. This is akin to
saying that indirect $CP$-violation does not occur among commuting states.

A weaker constraint comes from only considering negative 
$(mass)^2$ as nonphysical; this forbids certain ranges of values for
the $CP$-violating parameter.
For example, in the case when $CP$ is violated while $CPT$ is
preserved, this yields an upper bound for $\vert \epsilon\vert$.

\section{Conclusion. Perspective}
\label{section:perspective}

We have studied $CP$ violation for the neutral kaon system and  for
electroweak Higgs-like doublets, emphasizing their analogy,
and extended it to all possible values of the $CP$ violating parameter.

Our attention was drawn to ambiguities that arise in the spectrum of states
when the $CP$-odd commutator $[K^0,\ol{K^0}]$ vanishes.
The mass spectrum turns out to heavily depend
on the basis and on the $CP$-violating pattern attached to it.

This can look an academic problem since, in the real world and, in
particular, when electroweak interactions are turned on, such a commutator
is not expected to vanish.
However, adding to a Lagrangian, which is diagonal in a basis of $CP$
eigenstates, a term
$\kappa^2{\mathfrak C}$, where $\kappa^2$ is a function of the masses and
of the $CP$-violating parameter, alters the starting hierarchy in a way that
does not depend on the precise value of
$\mathfrak C$. In this framework, for small enough $\mathfrak C$,
a small perturbation is not excluded to trigger large hierarchies.

In particular, for certain ranges of values of the $CP$-violating parameter,
a negative $(mass)^2$ can occur in the  basis of $CP$ violating states, and
the theory becomes unstable.

We saw that there exist cases when the commutator always vanishes, which
maintains an ambiguity.  The  Higgs boson might fall into
this framework.  This needs a special investigation.

This work also suggests that discrete symmetries have to be handled with
care when reducing the number of degrees of freedom or constraining the
couplings in the Lagrangian. Some effects can be overlooked which play, in
particular, a role  in determining, even at the classical level,
the vacuum structure of the theory.

Our point of view has been different from  other studies in that we
did not investigate the origin of indirect $CP$-violation.
In particular, we paid no special attention to the potential that is
introduced in the Lagrangian.

This simple study  concerns the case where
only two particles or multiplets  are ``rotated''.
Since  one is free to perform a change of basis for any pair among the set of
Higgs-like doublets,  various hierarchies  can be expected
among these multiplets (see also section \ref{section:mh}).
This is left for a subsequent work \cite{machet3}.
 
\newpage\null
\appendix

\section{Dependent versus independent states}
\label{appendix:indep}

All relations written below for the kaon fields $\varphi$ of section
\ref{section:simple} are also true for the Higgs-like multiplets $\phi$ of
the next sections.

All relations written with charge conjugate fields are also valid with
their complex conjugates.

\subsection{ The case of a unitary change of basis}
\label{subsection:appu}

We come back to subsection \ref{subsection:unitary}
and examine some combinations
of fields relevant for writing the kinetic terms.

From the definition (\ref{eq:ls}), it is trivial to calculate,
\begin{equation}
\varphi_3^2 - \varphi_4^2 = (c_\theta^2 - s_\theta^2)
(\varphi_L^2 - \varphi_S^2) + 2s_\theta c_\theta(\varphi_L\varphi_S +
\varphi_S\varphi_L),
\label{eq:a1}
\end{equation}
and
\begin{equation}
\ol{\varphi_3}\varphi_3 + \ol{\varphi_4}\varphi_4 =
       \ol{\varphi_L}\varphi_L + \ol{\varphi_S}\varphi_S.
\label{eq:a2}
\end{equation}
The  l.h.s.'s of (\ref{eq:a1}) and (\ref{eq:a2}) are identical if one uses
the $C$ conjugation properties
$\ol{\varphi_3} = \varphi_3$ and $\ol{\varphi_4} = -\varphi_4$;
the two r.h.s. are also identical if one uses the relations (\ref{eq:genkk})
 between
$(\varphi_S, \ol{\varphi_S}, \varphi_L, \ol{\varphi_L})$.
The only difference is that (\ref{eq:a1})
is written with independent fields, while (\ref{eq:a2}) is written
with the two over-complete sets $(\varphi_3, \ol{\varphi_3}, \varphi_4,
\ol{\varphi_4})$ and $(\varphi_S, \ol{\varphi_S}, \varphi_L,
\ol{\varphi_L})$.

The form (\ref{eq:a2}), in which hermiticity is manifest, is the one that we
used, in particular, to write the kinetic terms.

One has also
\begin{equation}
\ol{\varphi_3}\varphi_4 + \ol{\varphi_4}\varphi_3 =
2s_\theta c_\theta (\ol{\varphi_L}\varphi_L - \ol{\varphi_S}\varphi_S)
-(c_\theta^2 - s_\theta^2)(\ol{\varphi_L}\varphi_S +
\ol{\varphi_S}\varphi_L)
\label{eq:a3}
\end{equation}
and
\begin{equation}
[\varphi_3,\varphi_4] = [\varphi_S,\varphi_L].
\label{eq:a4}
\end{equation}
%
\subsection{ The case of a non-unitary transformation}
\label{subsection:appnu}

 In the same way, for  subsection \ref{subsection:nonunit},
one has the relations
\begin{equation}
\varphi_3^2 - \varphi_4^2 = (\cosh^2\epsilon + \sinh^2\epsilon)( K_L^2 - K_S^2),
\label{eq:a5}
\end{equation}
and

\vbox{
\begin{eqnarray}
\ol{\varphi_3}\varphi_3 + \ol{\varphi_4}\varphi_4 &=& (\cosh^2\epsilon +
\sinh^2\epsilon)
\left((\cosh^2\epsilon + \sinh^2\epsilon)(\ol{K_L}K_L + \ol{K_S}K_S)\right.\cr
   && \hskip 3cm \left.-2\sinh\epsilon\cosh\epsilon(\ol{K_L}K_S +
\ol{K_S}K_L)\right),
\label{eq:a6}
\end{eqnarray}
}

and
\begin{equation}
\ol{K_L}K_L + \ol{K_S}K_S =
(\ol{\varphi_3}\varphi_3 + \ol{\varphi_4}\varphi_4)
+\frac{2\sinh\epsilon\cosh\epsilon}{\cosh^2\epsilon + \sinh^2\epsilon}
(\ol{\varphi_3}\varphi_4 + \ol{\varphi_4}\varphi_3).
\label{eq:a7}
\end{equation}
One has also

\vbox{
\begin{eqnarray}
\ol{\varphi_3}\varphi_4 + \ol{\varphi_4}\varphi_3 &=& 
(\cosh^2\epsilon + \sinh^2\epsilon) \left(
-2\sinh\epsilon\cosh\epsilon(\ol{K_L}K_L + \ol{K_S}K_S)\right.\cr
&& \left. \hskip 2cm +
(\cosh^2\epsilon + \sinh^2\epsilon)(\ol{K_L}K_S + \ol{K_S}K_L)\right),
\label{eq:a8}
\end{eqnarray}
}

and
\begin{equation}
[\varphi_3,\varphi_4] = (\sinh^2\epsilon +
\cosh^2\epsilon)[K_L(\epsilon),K_S(\epsilon)].
\label{eq:a9}
\end{equation}

This shows again that the choice of the basis is important and that working
with non-independent states is ambiguous.

\newpage\null
\begin{em}

\end{em}
%
%

\begin{thebibliography}{50}
%
\bibitem{BrancoLavouraSilva}
        G.C. BRANCO, L. LAVOURA \& J.P. SILVA: ``$CP$ violation'' (Oxford
University Press, 1999), and references therein.

\bibitem{Belusevic}
        R. BELU\u{S}EVI\'C: ``Neutral kaons'' (Springer Tracts in Modern
Physics, vol. 153, 1999), and references therein.

\bibitem{Lee}
       T.D. LEE: ``A Theory of Spontaneous $T$ Violation'', Phys. Rev. D
                   8 (1973) 1226.

\bibitem{Weinberg}
        S. WEINBERG: ``Gauge Theory of $CP$ Nonconservation'', Phys. Rev.
Lett. 37 (1976) 657.

\bibitem{GunionHaberKaneDawson}
        J.F. GUNION, H.E. HABER, G. KANE \& S. DAWSON: ``The Higgs Hunter's
Guide'' (Addison Wesley, Frontiers in Physics, 1990).

\bibitem{Lavoura1}
       L. LAVOURA: ``Models of $CP$ violation exclusively via
           neutral-scalar exchange'', IJMP A 9 (1994) 1873-1888.

\bibitem{Branco}
       G.C. BRANCO: ``Spontaneous $CP$ nonconservation and neutral flavor
              conservation: A minimal model'', Phys. Rev. D 22 (1980) 2901.

\bibitem{LavouraSilva}
       L. LAVOURA \& J.P. SILVA: ``Fundamental $CP$-violating quantities in
             a $SU(2) \times U(1)$ model with many Higgs doublets'',
             hep-ph/9404276, Phys. Rev. D 50 (1994) 4619-4624.

\bibitem{CabibboKobayashiMaskawa}
       N. CABIBBO: ``Unitary symmetry and leptonic decays'',
                           Phys. Rev. Lett. 10 (1963) 531;\l
       M. KOBAYASHI and T. MASKAWA:  ``$CP$-Violation in the Renormalizable
          Theory of Weak Interactions'', Prog. Theor. Phys. 49 (1973) 652.

\bibitem{GlashowWeinberg}
       S.L. GLASHOW \& S. WEINBERG: ``Natural conservation laws for neutral
              currents'', Phys. Rev. D 15 (1977) 1958.

\bibitem{Georgi}
       H. GEORGI: ``A model of soft $CP$ violation'', Hadronic J. 1 (1978)
155.

\bibitem{MendezPomarol}
       A. M\'ENDEZ \& A. POMAROL: ``Signals of $CP$ violation in the Higgs
sector'', Phys. Lett. B 272 (1991) 313.

\bibitem{Commins}
       E.D. COMMINS \& P.H. BUCKSBAUM: ``Weak interactions of leptons and
                 quarks'' (Cambrige University Press, 1983).

\bibitem{Pokorsky}
       see for example:\l
       S. POKORSKY: ``Gauge Field Theories'' (Cambridge Monographs on
Mathematical Physics, 1999), p.~58-60.

\bibitem{CPLEAR}
      CPLEAR Collaboration (A. Angelopoulos et al.):
``First direct observation of time-reversal non-invariance in the neutral
kaon system''  CERN-EP-98-153, Phys.Lett.B 444 (1998) 43-51.

\bibitem{AlvarezGaume}
       L. ALVAREZ-GAUM\'E, C. KOUNNAS, S. LOLA \& P. PAVLOPOULOS:
      ``Violation of Time-Reversal Invariance and CPLEAR measurements'',
       Phys. Lett. B 458 (1999)347-354, hep-ph/9812326.

\bibitem{GlashowSalamWeinberg}
      S.L. GLASHOW: ``Partial-symmetry of weak interactions'',
                                Nucl. Phys. 22 (1961) 579;\l
      A. SALAM: ``Weak and electromagnetic interactions'',
             in ``Elementary Particle Theory: Relativistic Groups and
             Analyticity'' (Nobel symposium No 8), edited by N. Svartholm
             (Almquist and Wiksell, Stockholm 1968);\l
      S. WEINBERG: ``A model of leptons'', Phys. Rev. Lett. 19 (1967) 1264.

\bibitem{Machet1}
      B. MACHET: ``Chiral scalar fields, custodial symmetry in electroweak
         $SU(2)_L \times U(1)$ and the quantization of the electric charge'',
          Phys. Lett. B 385 (1996) 198-208, hep-ph/9606239.

\bibitem{Machet2}
       B. MACHET: ``Indirect $CP$ violation in an electroweak $SU(2)_L
            \times U(1)$ gauge theory of chiral mesons'', hep-ph/9804417.

\bibitem{GeorgiNanopoulos}
        H. GEORGI \& D.V. NANOPOULOS: ``Suppression of flavour changing
            effects from neutral spinless meson exchange in gauge theories'',
            Phys. Lett. 82 B (1979) 95.

\bibitem{AhluwaliaKirchbach}
       D.V. AHLUWALIA \& M. KIRCHBACH: ``$(1/2,1/2)$ representation space;
             an ab initio construct'', hep-th/0101009.

\bibitem{Wolfenstein}
       L. WOLFENSTEIN: ``Parametrization of the Kobayashi-Maskawa matrix'',
 Phys. Rev. Lett. 51 (1983) 1945.

\bibitem{machet3}
       B. MACHET: ``Hierarchies of masses for scalar / pseudoscalar fields'',
in preparation.
%
\end{thebibliography}
\end{document}